\documentclass[journal,english,twocolumn]{IEEEtran}

\usepackage[T1]{fontenc}
\usepackage{amsmath}
\usepackage{graphicx}
\usepackage{amssymb}
\usepackage{subfigure}
\usepackage{epstopdf}
\usepackage{gensymb}
\usepackage{cite}
\usepackage[dvipsnames]{xcolor}

\makeatletter
\usepackage[font=footnotesize]{caption}
\makeatother

\usepackage{tikz}
\newcommand\copyrighttext{%
  \footnotesize \textcopyright 2018 IEEE. Personal use of this material is permitted.
  Permission from IEEE must be obtained for all other uses, in any current or future 
  media, including reprinting/republishing this material for advertising or promotional 
  purposes, creating new collective works, for resale or redistribution to servers or 
  lists, or reuse of any copyrighted component of this work in other works. }
\newcommand\copyrightnotice{%
\begin{tikzpicture}[remember picture,overlay]
\node[anchor=south,yshift=0pt] at (current page.south) {\fbox{\parbox{\dimexpr\textwidth-\fboxsep-\fboxrule\relax}{\copyrighttext}}};
\end{tikzpicture}%
}

\usepackage{babel}

\usepackage[final]{changes} 

\colorlet{Changes@Color}{blue}
\begin{document}
\bstctlcite{IEEEexample:BSTcontrol}

\title{Cold-Source Noise Measurement of a Differential Input Single-Ended Output Low-Noise Amplifier Connected to a Low-Frequency Radio Astronomy Antenna}

\author{Adrian Sutinjo \IEEEmembership{Senior Member,~IEEE}, Daniel Ung \IEEEmembership{Graduate Student Member,~IEEE}, and Budi Juswardy
\thanks{\textbf{IEEE Trans. Antennas. Propagat., accepted, 24 June 2018. The authors are with the International Centre for Radio Astronomy Research (ICRAR)/Curtin University.}}
}

\maketitle
\copyrightnotice

\begin{abstract}
We present two methods for measuring the noise temperature of a differential input single-ended output (DISO) Low-Noise Amplifier (LNA) connected to an antenna. The first method is direct measurement of the DISO LNA and antenna in an anechoic chamber at ambient temperature. The second is a simple and low-cost noise parameter extraction of the DISO device using a coaxial long cable. The reconstruction of the DISO noise parameter from the noise wave measurements of the DISO LNA with one terminated input port is discussed in detail.  We successfully applied these methods to the Murchison Widefield Array LNA and antenna.
\end{abstract}


\thispagestyle{empty}

\section{Introduction}
\label{sec:intro}
Differential input single-ended output (DISO) Low-Noise Amplifiers (LNAs) are widely used in low-frequency radio telescopes, for example the Murchison Widefield Array (MWA)~\cite{Lonsdale_2009, 2013PASA...30....7T}, the Low-Frequency Array (LOFAR)~\cite{GHT_00, 2013A&A...556A...2V}, Low-Frequency Square Kilometer Array (SKA-Low) prototypes~\cite{Eloy_EXAP2015, 7293140}, the Long Wavelength Array (LWA)~\cite{Bradley_LWA_2005, Paravastu_LWA_2007, LWA1_6420880} and the Giant Ukrainian Radio Telescope (GURT)~\cite{6379725}. In these examples, the differential input of the LNAs are connected directly to a balanced antenna which is typically a dipole-type antenna. Particularly at the low end of the frequency band, these antennas are strongly mismatched to standard $50~\Omega$ (or $100~\Omega$ for a differential port) offered by a typical measurement instrument. Hence, the noise temperature of the LNA ($T_{\mathrm{LNA}}$) connected to the antenna may differ significantly from $T_{\mathrm{LNA}}$ to a matched source and must be accounted for appropriately.

There are two common approaches to making this measurement. One way is to measure the LNA connected to the antenna using sky noise and ambient noise from absorbers as hot/cold sources~\cite{6058624, 6902591}. A related approach is use room temperature load as hot source and cooled load as cold source~\cite{6328666}. The other is to use knowledge of LNA noise parameters and the antenna reflection coefficient ($\Gamma_\mathrm{AUT}$) to calculate the noise temperature, $T_{\mathrm{LNA}}(\Gamma_s)$~\cite{Leo_16_7506352, Hu_04_1295149, Wedge_92_168757}. However, these standard approaches are not easily applicable to low-frequency radio astronomy (tens to hundreds of MHz). The hot/cold approach works well at microwave frequencies because relatively narrow beamwidth can be achieved with a reasonable AUT size. This is not the case at low frequencies. Furthermore, in the frequency range of tens to hundreds of MHz, the average sky noise ($T_{sky}\approx60\lambda^{2.55}$~K) transitions from being hotter to cooler than ambient temperature at about 160~MHz such that hot/cold contrasts are difficult to discern around this frequency. 

The noise parameters approach often suffers from lack of noise parameter information which could be cost-prohibitive to obtain. For example, in the case of the MWA (80-300~MHz), the LNA data sheet does not provide noise parameters below 500 MHz~\cite{ATF-54143}. Although commercial-off-the-shelf impedance tuners are available for the frequency of interest, they are costly and sometimes multiple tuners have to be purchased to cover the entire frequency range. In addition, the impedance tuners are single-ended devices while the DISO LNA has two inputs. It is desirable to obtain the noise parameters of the LNA in its DISO configuration directly as opposed to combining two single-ended input single-ended output (SISO) LNAs in simulation. 

This paper discusses a few contributions to overcome these challenges. We address both direct measurement and noise parameter extraction. For direct measurement, we demonstrate an intuitive measurement method that involves placing the antenna connected to the LNA in an anechoic chamber acting as an ambient (cold) noise source. Regarding noise parameter extraction of a DISO device, we discuss an extraction technique using a cold long cable without an input balun and a method to combine the measurement results to correctly reconstruct the DISO noise parameters. The results will be demonstrated with measurement of an MWA DISO LNA.  

Finally, we review our contributions in comparison to existing literature. The theory for multiport noise parameter extraction has been discussed in~\cite{7824361, 7577488} and demonstrated using a passive four-port network. This paper provides an example and gives simple theory for noise parameter extraction of an important class of 3-port active devices (i.e., DISO LNA). Noise parameter extraction using long cables has been used in low-frequency radio astronomy cosmology~\cite{RDS:RDS5996, PAS:9560390}, however, the LNAs in question are SISO devices. Furthermore, we address in detail the question of proper selection of the long cable. Ideal calculation of DISO noise parameters using known constituent SISO noise parameters is discussed in~\cite{Prinsloo_14_6813592}. We assume no prior knowledge of SISO noise parameters and the reconstruction of the DISO noise parameters relies purely on \emph{measured} quantities using the actual DISO LNA board. The immediate context of this paper is noise temperature of a single DISO LNA connected to one antenna. Hence, noise coupling that occurs in a low-frequency radio astronomy arrays~\cite{5725170, 7079488} is not reflected in the results. However, the direct measurement method we discuss may be applied to an antenna array if it is placed in a sufficiently large anechoic chamber. In addition, the noise parameter extraction method and the results thereof are useful for noise coupling calculation in an array environment. 

This paper is organized as follows. Sec.~\ref{sec:direct} and Sec.~\ref{sec:extract} provide a high-level overview of the direct measurement and the DISO noise parameter extraction techniques. Sec.~\ref{sec:meas_res} discusses specific considerations for successful measurement and presents measurement results. Uncertainty estimates for both methods are given in Sec.~\ref{sec:uncertainty}. Sec.~\ref{sec:concl} presents our conclusions. 

\section{Direct Measurement}
\label{sec:direct}
\begin{figure}[htb]
	\begin{center}
	\includegraphics[width=2.75in]{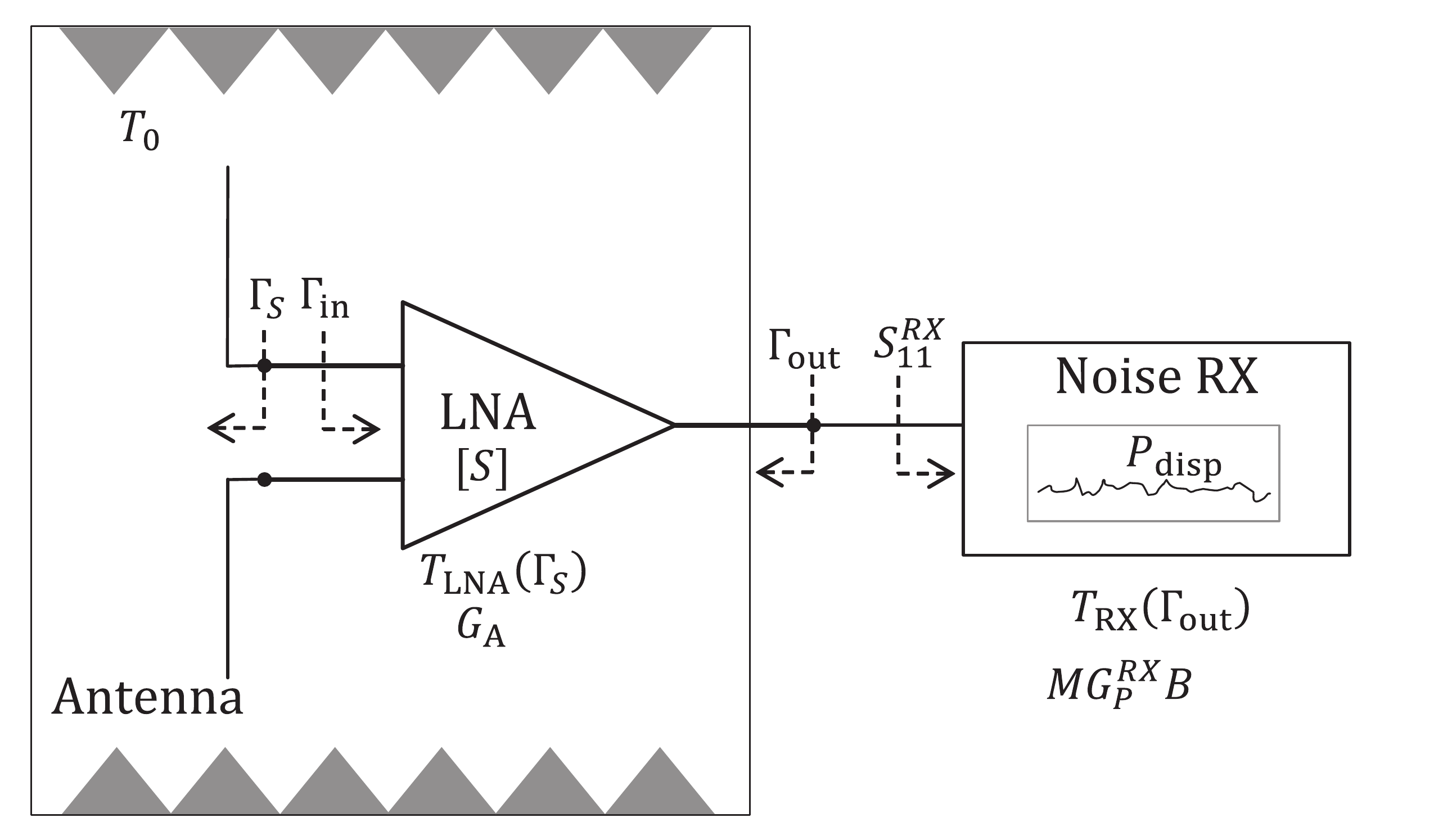}
	\end{center}
\caption{A depiction of direct measurement of a DISO LNA connected to an antenna by placing them in an anechoic chamber. $P_\mathrm{disp}$ is the displayed noise power, $M$ is the mismatch loss at the input of the noise receiver (RX), $G_{P}^{RX}B$ is the power gain-bandwidth product of the RX, $G_{A}$ is the available gain of the LNA~\cite{Gonzalez_1997_ch3}, $[S]=[S_{dd11},S_{ds12}; S_{sd21},S_{ss22}] $ is the differential to single-ended $S$-parameters of the LNA; $\Gamma_\mathrm{in}$ and $\Gamma_\mathrm{out}$ are the input and output reflection coefficients~\cite{Gonzalez_1997_ch3}, respectively.}
\label{fig:direct_diag}
\end{figure}

\begin{figure}[htb]
	\begin{center}
	\includegraphics[width=2.75in]{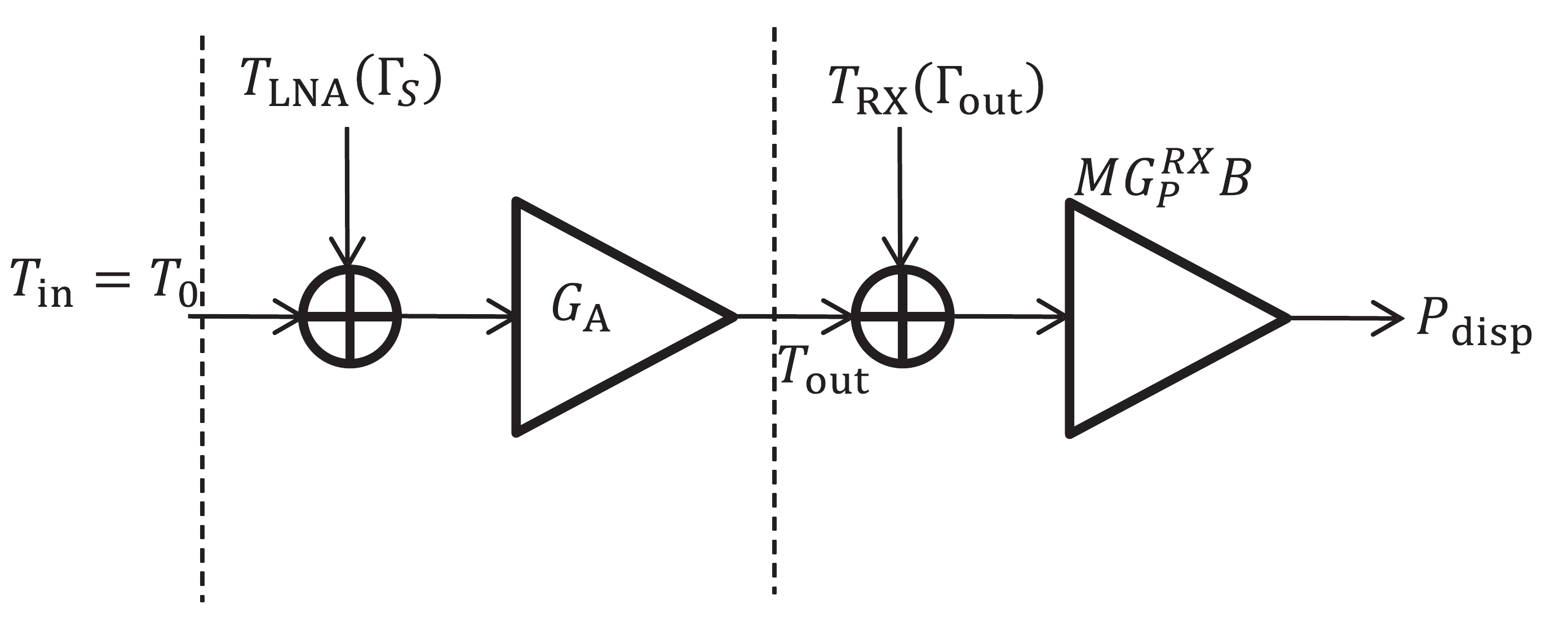}
	\end{center}
\caption{The noise flow diagram of the direct measurement setup. $T_\mathrm{out}$ is the total noise power density at the output of the LNA for available noise power density $T_\mathrm{in}$ at the input and  source reflection coefficient $\Gamma_s$.}
\label{fig:noise_chn}
\end{figure}

This method is inspired by the ``cold-source'' noise figure measurement technique employed by Keysight Precision Network Analyzer (PNA-X)~\cite{Keysight_PNAX}. Here, we modify the technique to  accept the ambient noise temperature emitted by the anechoic chamber as cold noise source as shown in Fig.~\ref{fig:direct_diag}. Also, this measurement involves differential to single-ended $S$-parameters. The noise flow diagram is given in Fig.~\ref{fig:noise_chn} to clarify the measurement strategy.   

In Figs.~\ref{fig:direct_diag} and~\ref{fig:noise_chn}, noise measurement is performed by the noise receiver (RX) which displays the noise power, $P_\mathrm{disp}$. The quantity of interest is $T_{\mathrm{LNA}}(\Gamma_s)$. The available input noise is the ambient noise temperature, $T_\mathrm{in}=T_0$. Fig.~\ref{fig:noise_chn} suggests
\begin{eqnarray}
P_\mathrm{disp} &=& k \left[T_\mathrm{out} + T_{RX}(\Gamma_\mathrm{out}) \right] MG_{P}^{RX}B \nonumber \\
T_\mathrm{out}&=&G_{A}\left[T_\mathrm{in}+T_\mathrm{LNA}(\Gamma_s)\right]
\label{eqn:direct_meas_eqns}
\end{eqnarray}
\noindent 
where $k$ is the Boltzmann constant. After rearranging, it can shown be that
\begin{eqnarray}
T_\mathrm{out}&=&\frac{P_\mathrm{disp}(\Gamma_s)}{kMG_{P}^{RX}B}-T_{RX}(\Gamma_\mathrm{out}) \nonumber \\
T_\mathrm{LNA}(\Gamma_s)&=&\frac{T_\mathrm{out}}{G_{A}}-T_0
\label{eqn:T_LNA_ant}
\end{eqnarray}
To find $T_\mathrm{LNA}(\Gamma_s)$, we need to obtain the available gain of the LNA
\begin{eqnarray}
G_{A}=\frac{1-|\Gamma_s|^2}{|1-S_{dd11}\Gamma_s|^2}\frac{|S_{sd21}|^2}{1-|\Gamma_\mathrm{out}|^2}, 
\label{eqn:GA}
\end{eqnarray}
which requires the measurement of reflection coefficient of the antenna under test (AUT) where $\Gamma_s=\Gamma_{\mathrm{AUT}}$ and the two differential to single-ended $S$-parameters~\cite{392911, Fan_03_1271579} shown above. We also need the mismatch factor of the RX
\begin{eqnarray}
M=\frac{\left(1-\left|\Gamma_\mathrm{out} \right|^2 \right) \left(1-\left|S_{11}^{RX} \right|^2 \right)}{\left|1-S_{11}^{RX}\Gamma_\mathrm{out}\right|^2}, 
\label{eqn:ML}
\end{eqnarray}
which requires measurement of the input reflection coefficient of the RX, $S_{11}^{RX}$. In addition, we require the power gain-bandwidth product $G_{P}^{RX}B$ and the noise temperature of RX $T_{RX}$ given that  it sees $\Gamma_\mathrm{out}$ at its input. We obtain the latter two quantities through RX calibration described next. 

\subsection{Noise Receiver Calibration}
\label{sec:RX cal}

\begin{figure}[htb]
	\begin{center}
	\includegraphics[width=2.75in]{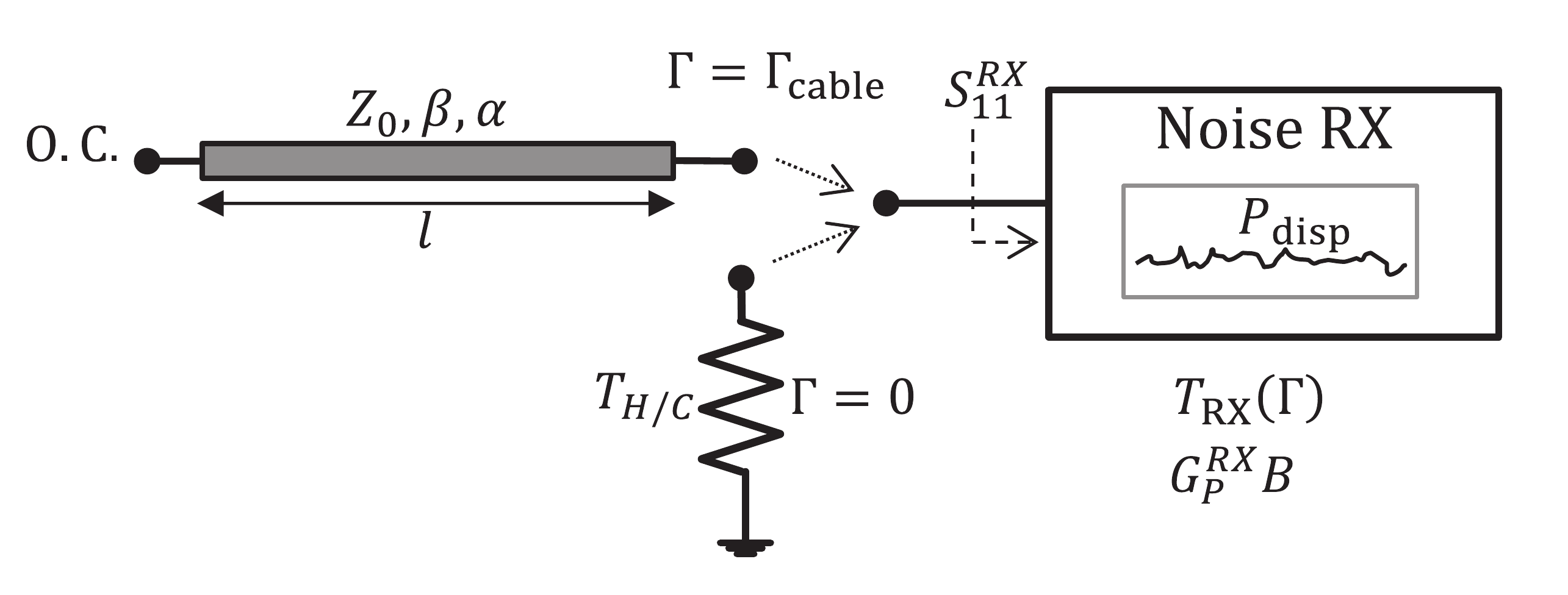}
	\end{center}
\caption{Noise receiver calibration. The hot/cold noise source is used to measure $G_{P}^{RX}B$. The lossy open circuited (o.c.) coax cable is used present a mismatch source with available noise at ambient temperature.  }
\label{fig:noise_cal}
\end{figure}

Noise receiver calibration consists of two steps: measuring power gain-bandwidth of the RX and extraction of noise parameters of the RX. 

\subsubsection{Power Gain-Bandwidth Measurement}
\label{sec:gain_bw}

The power gain-bandwidth is obtained using a matched ($\Gamma=0$) hot/cold noise source as follows. Connect a hot/cold source to the input of the RX, then measure
\begin{eqnarray}
P_\mathrm{disp}^{C}=k\left[T_{C} + T_{RX}(\Gamma=0)\right]\left(1-\left|S_{11}^{RX} \right|^2 \right)G_{P}^{RX}B \nonumber \\
P_\mathrm{disp}^{H}=k\left[T_{H} + T_{RX}(\Gamma=0)\right]\left(1-\left|S_{11}^{RX} \right|^2 \right)G_{P}^{RX} B
\label{fig:H_C cal}
\end{eqnarray}
where $(1-\left|S_{11}^{RX} \right|^2)$ is the only remaining term in \eqref{eqn:ML} for $\Gamma=0$. Taking the difference, we get
\begin{eqnarray}
P_\mathrm{disp}^{H}-P_\mathrm{disp}^{C}=k\left(1-\left|S_{11}^{RX}\right|^2\right)G_{P}^{RX}B\left(T_{H}-T_{C}\right)
\label{fig:H_C cal_diff}
\end{eqnarray}
The excess noise ratio, $\mathrm{ENR}=(T_{H}-T_{C})/T_{0}$, is assumed known \textit{a priori}. This allows us to write
\begin{eqnarray}
G_{P}^{RX}B=\frac{P_\mathrm{disp}^{H}-P_\mathrm{disp}^{C}}{kT_{0}\left(1-\left|S_{11}^{RX}\right|^2\right)\mathrm{ENR}}
\label{eqn:GpB}
\end{eqnarray}
$G_{P}^{RX}$ is the power gain of the RX that relates the power injected to its input to the displayed power.

\subsubsection{RX Noise Parameter Extraction}
\label{sec:RX_noise_par}
For any ambient temperature noise source, the knowledge of $G_{P}^{RX} B$ enables us to infer $T_{RX}$ for arbitrary $\Gamma$ by measuring $P_\mathrm{disp}$.
\begin{eqnarray}
P_\mathrm{disp}(\Gamma)=k\left[T_{0} + T_{RX}(\Gamma)\right]M'G_{P}^{RX} B
\label{fig:P_disp_meas}
\end{eqnarray}
where $M'$ is \eqref{eqn:ML} with $\Gamma_\mathrm{out}$ replaced with $\Gamma$. The receiver noise temperature is
\begin{eqnarray}
T_{RX}(\Gamma)=\frac{P_\mathrm{disp}(\Gamma)}{\mathrm{M'}kG_{P}^{RX}B}-T_{0}
\label{eqn:T_RX_meas}
\end{eqnarray}
To extract the noise parameters ($T_{RX}^\mathrm{min}, \Gamma_{RX}^\mathrm{opt}, N_{RX}$) of the RX, we present a number of $\Gamma_n$'s using an open circuited long cable and a matched cold source. The resulting $T_{RX}$ for a given $\Gamma_{n}$ is given by
\begin{eqnarray}
T_{RX}(\Gamma_n) = T_{RX}^\mathrm{min}+4T_0N_{RX}\frac{|\Gamma_n-\Gamma_{RX}^\mathrm{opt}|^2}{(1-|\Gamma_n|^2)(1-|\Gamma_\mathrm{opt}|^2)}
\label{eqn:T_RX_n}
\end{eqnarray}
This measurement technique is similar to~\cite{Hu_04_1295149} except for the use of ambient temperature cable~\cite{RDS:RDS5996} as a noise source. We will discuss the details of this step in Sec.~\ref{sec:meas_res}. For now, it suffices to say that the RX noise parameters may be extracted by measuring $T_{RX}(\Gamma_n)$ for four or more $\Gamma_n$'s. 

\subsection{Summary of the Direct Measurement Method}
\label{sec:sum_direct}
\begin{enumerate}
\item Calibrate the RX. Obtain power gain-bandwidth product, $G_{P}^{RX}B$; measure $S_{11}^{RX}$ and then obtain noise parameters: $T_{RX}^\mathrm{min}, \Gamma_{RX}^\mathrm{opt}, N_{RX}$. We can now  calculate $T_{RX}$ for arbitrary output reflection coefficient $\Gamma_\mathrm{out}$ of the LNA under test.
\item  Measure $|S_{sd21}|$, $S_{dd11}$ of the LNA. 
\item  Connect the LNA to the antenna in the anechoic chamber and measure $\Gamma_\mathrm{out}$ of the LNA connected to the antenna.
\item Measure $P_\mathrm{disp}$ with the antenna connected to the LNA.
\item Measure the antenna reflection coefficient in the anechoic chamber, $\Gamma_{\mathrm{AUT}}$. Obtain \eqref{eqn:GA} and \eqref{eqn:ML}.
\item Calculate $T_\mathrm{LNA}(\Gamma_{\mathrm{AUT}})$ using \eqref{eqn:T_LNA_ant}.
\end{enumerate}

\section{DISO Noise Parameter Extraction}
\label{sec:extract}

\begin{figure}[htb]
	\begin{center}
	\includegraphics[width=2.5in]{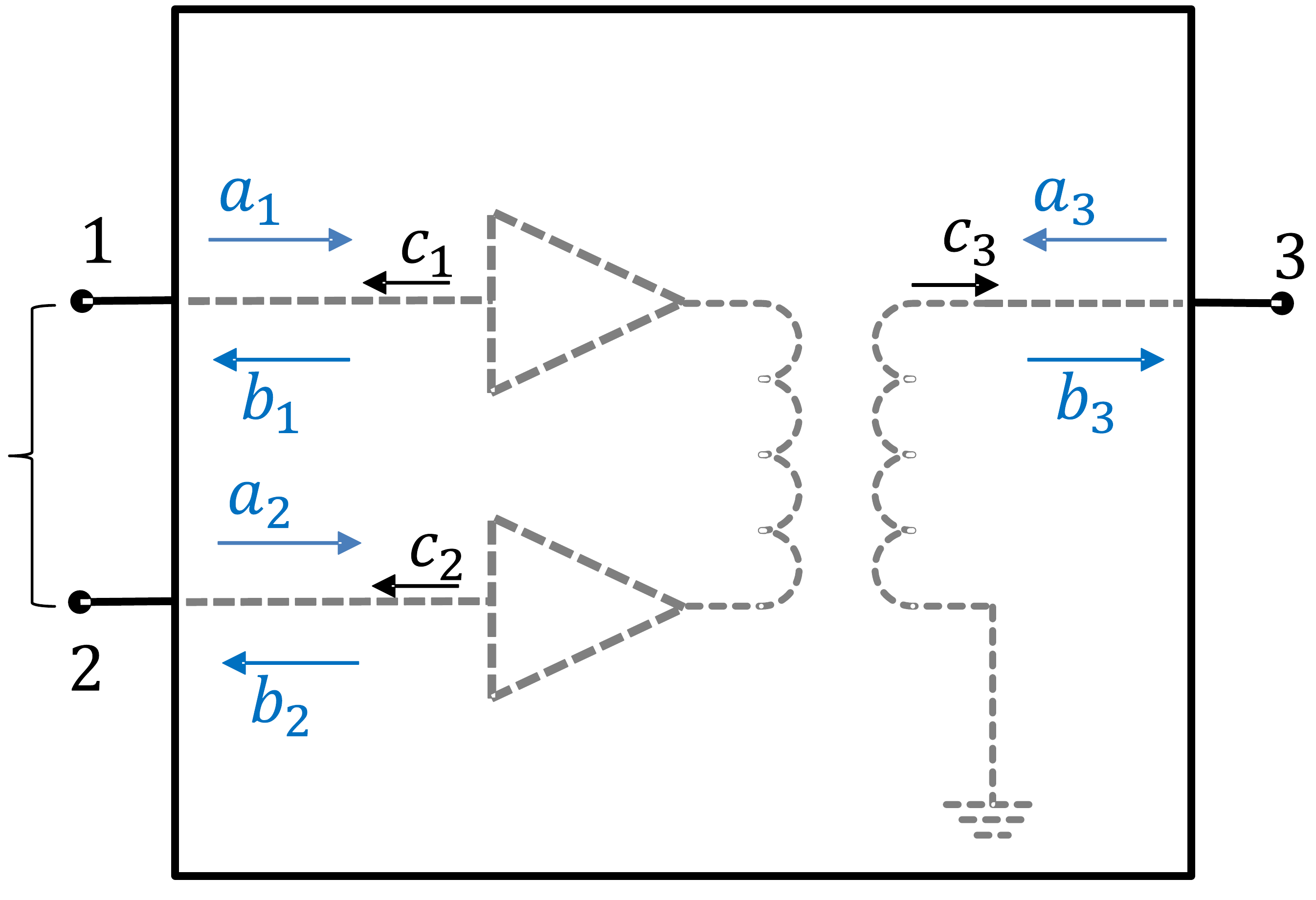}
	\end{center}
\caption{The three-port device representing a nominal DISO device. In the final product, ports~1 and~2 are excited as a differential port. $a_{i}$ and $b_{i}$ are the incident and reflective waves, respectively. $c_{i}$ is the intrinsic noise wave ~\cite{Wedge_92_168757, Hu_04_1295149,Leo_16_7506352}.}
\label{fig:3port_dev}
\end{figure}

\subsection{Multiport Noise Figure and Parameters}
\label{sec:multi}
We apply the multiport noise factor expression as a function of port reflection coefficient and the correlation matrix of the incident noise waves~\cite{Randa_01_954781} to the network in Fig.~\ref{fig:3port_dev}. The noise factor expression for output port~3 is: 
\begin{eqnarray}
&&F_{3}\left(\mathbf{\Gamma},\mathbf{A}\right) =\nonumber \\
&&
1+
\frac{\left[\left(\mathbf{I}-\mathbf{S}\mathbf{\Gamma}\right)^{-1} \hat{\mathbf{N}}\left(\mathbf{I}-\mathbf{S}\mathbf{\Gamma}\right)^{-1H}\right]_{3,3}}
{kT_{0}\left[\left(\mathbf{I}-\mathbf{S}\mathbf{\Gamma}\right)^{-1} \mathbf{SAS}^{H}\left(\mathbf{I}-\mathbf{S}\mathbf{\Gamma}\right)^{-1H}\right]_{3,3}}
\label{eqn:Multiport_F}
\end{eqnarray}
where we are using the single-ended port numbering convention for now and $(.)^{H}$ is the conjugate transpose operation. However, \eqref{eqn:Multiport_F} is sufficiently general to handle the single-ended to differential port conversion as we will see later.
\begin{eqnarray}
\hat{\mathbf{N}}=
\left( \begin{array}{ccc}
\left< \left|c_{1}\right|^2 \right> & \left< c_{1}c_{2}^* \right> & \left< c_{1}c_{3}^* \right> \\
\left< c_{2}c_{1}^* \right>  & \left< \left|c_{2}\right|^2 \right> & \left< c_{2}c_{3}^* \right> \\
\left< c_{3}c_{1}^* \right>  & \left< c_{3}c_{2}^{*} \right>  & \left< \left|c_{3}\right|^2 \right>
\end{array} \right) 
\label{eqn:Noise_corr}
\end{eqnarray}
 is the intrinsic noise correlation matrix. This is the \emph{unknown} matrix of interest (with 9 unknowns).
$\mathbf{S}$ is the $S$-parameter matrix obtained through vector network analyzer (VNA) measurements.
$\mathbf{A}$ is the incident noise correlation matrix due to thermal noise of the terminations relative to $kT_{0}$ and $\mathbf{\Gamma}$ is the source reflection coefficient matrix. In-depth discussion of $\mathbf{A}$ and $\mathbf{\Gamma}$ is given in~\cite{Randa_01_954781}; however, this is not critical for our current purpose because, for a DISO device, these matrices take on simplified and intuitive forms as we will discuss in Sec.~\ref{sec:apply}. Subscript $._{3,3}$ refers to the entry at the 3rd row and 3rd column. 

The key here is to recognize that the noise parameters of a DISO device is fully described by \eqref{eqn:Noise_corr}. The question now is how to populate $\hat{\mathbf{N}}$ with as few measurements as possible.

\subsection{Three-Port to Two-Port Noise Wave Conversion}
\label{sec:three_2_conv}
The relationship between $S$~parameters, $a_{i}$, $b_{i}$, and $c_{i}$ in Fig.~\ref{fig:3port_dev} is given by~\cite{Wedge_92_168757}

\begin{eqnarray}
\left( \begin{array}{c}
b_{1} \\
b_{2} \\
b_{3} \\
\end{array} \right) 
&=&
\left( \begin{array}{ccc}
S_{11} & S_{12} & S_{13}\\
S_{21} & S_{22} & S_{23}\\
S_{31} & S_{32} & S_{33}\\
\end{array} \right) 
\left( \begin{array}{c}
a_{1} \\
a_{2} \\
a_{3} \\
\end{array} \right) 
+
\left( \begin{array}{c}
c_{1} \\
c_{2} \\
c_{3} \\
\end{array} \right) \nonumber \\
\mathbf{b}&=&\mathbf{S}\mathbf{a}+\mathbf{c}
\label{eqn:S_abc}
\end{eqnarray}
The 3-port $S$-parameters are known through measurement. We extract the noise parameters of an "effective" two-port device at a time, with the unused port terminated with a resistive load. These noise parameters are convertible to noise waves~\cite{Leo_16_7506352, Wedge_92_168757} and vice versa.

\begin{figure}[htb]
	\begin{center}
	\includegraphics[width=2.5in]{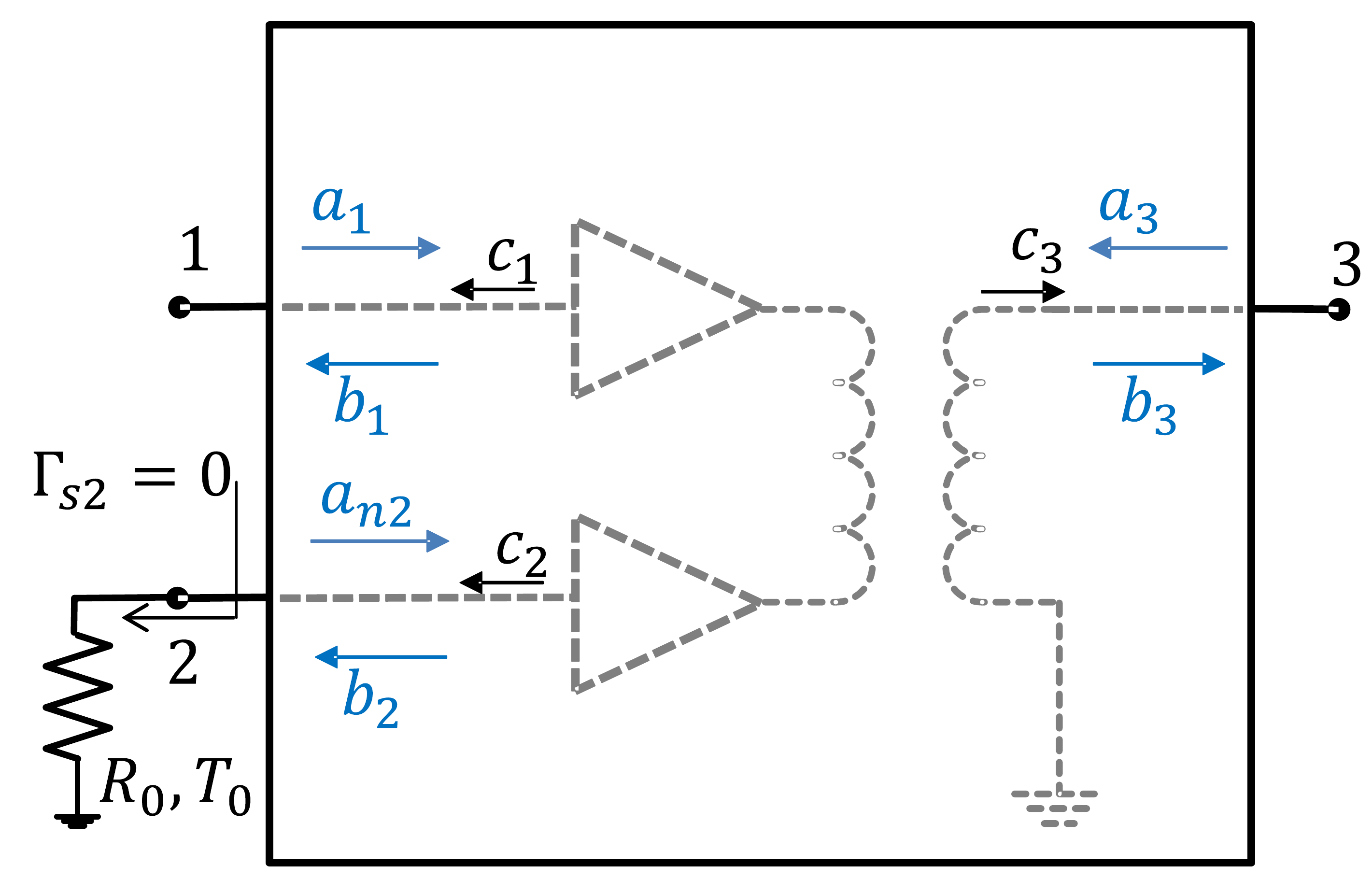}
	\end{center}
\caption{Three-port to two-port conversion through reflectionless termination of an unused port}
\label{fig:3_2_port_convert}
\end{figure}

\subsubsection{Terminating Port~2}
\label{sec:term_port2}
Consider terminating port~2 with a reflectionless load at ambient temperature as shown in Fig.~\ref{fig:3_2_port_convert} such that $a_{2}=a_{n2}$ (where $|a_{n2}|^{2}=kT_{0}$)~\cite{Grosch_93_245675}. Port~1 is connected to the open circuited long cable\footnote{We use the long cable as a low-cost option.  Alternatively, one could use a source impedance tuner.} (similar to Sec.~\ref{sec:RX_noise_par}) to extract the noise parameters of the effective two-port. Relating only ports~1 and~3, we obtain
\begin{eqnarray}
\left( \begin{array}{c}
b_{1} \\
b_{3} \\
\end{array} \right) 
&=&
\left( \begin{array}{c}
S_{11} a_{1}  + S_{13} a_{3}\\
S_{31} a_{1}  + S_{33} a_{3}\\
\end{array} \right) 
+ 
\left( \begin{array}{c}
S_{12} a_{n2} + c_{1}\\
S_{32} a_{n2} + c_{3}\\
\end{array} \right) \nonumber \\
&=&
\left( \begin{array}{cc}
S_{11} & S_{13} \\
S_{31} & S_{33} \\
\end{array} \right) 
\left( \begin{array}{c}
a_{1}\\
a_{3}\\
\end{array} \right) 
+ 
\left( \begin{array}{c}
c_{e1}\\
c_{e3}\\
\end{array} \right)
\label{eqn:terminate_port2_relate}
\end{eqnarray}

The intrinsic noise correlation matrix of the effective two-port is:
\begin{eqnarray}
\mathbf{C}_{e:1,3} =
\left( \begin{array}{cc}
\left< \left|c_{e1}\right|^2 \right>_{1,3}  & \left< c_{e1}c_{e3}^* \right>_{1,3} \\
\left< c_{e1}^*c_{e3} \right>_{1,3}  & \left< \left|c_{e3}\right|^2 \right>_{1,3} \\
\end{array} \right) 
\label{eqn:Ce_1_3}
\end{eqnarray}
where
\begin{eqnarray}
\left< \left|c_{e1}\right|^2 \right>_{1,3} &=& \left< \left|c_{1}\right|^2 \right> + \left|S_{12}\right|^2kT_{0} \label{eqn:Ce_1_sqr} \\
\left< \left|c_{e3}\right|^2 \right>_{1,3} &=& \left< \left|c_{3}\right|^2 \right> + \left|S_{32}\right|^2kT_{0} \label{eqn:Ce_3_sqr} \\
\left< c_{e1}c_{e3}^* \right>_{1,3} &=& \left< c_{1}c_{3}^* \right> + S_{12}S_{32}^* kT_{0} 
\label{eqn:Ce1_ce3}
\end{eqnarray}
The desired quantities are $\left< \left|c_{1}\right|^2 \right>$, $\left< \left|c_{3}\right|^2 \right>$, and $\left< c_{1}c_{3}^* \right>$ in the right-hand-side (RHS) of \eqref{eqn:Ce_1_sqr}--\eqref{eqn:Ce1_ce3} which are four out of the nine unknowns in~\eqref{eqn:Noise_corr}. These quantities are extracted from the measured noise correlation matrix of the effective two-port through knowledge of the $S$-parameters of the device under test (DUT) and the ambient temperature.

\subsubsection{Terminating Port~1}
\label{sec:term_port1}
Repeat the process above by terminating port~1 and connecting port~2 to the open circuited long cable. We obtain
\begin{eqnarray}
\mathbf{C}_{e:2,3} =
\left( \begin{array}{cc}
\left< \left|c_{e2}\right|^2 \right>_{2,3}  & \left< c_{e2}c_{e3}^* \right>_{2,3} \\
\left< c_{e2}^*c_{e3} \right>_{2,3}  & \left< \left|c_{e3}\right|^2 \right>_{2,3} \\
\end{array} \right) 
\label{eqn:Ce_2_3}
\end{eqnarray}
where
\begin{eqnarray}
\left< \left|c_{e2}\right|^2 \right>_{2,3} = \left< \left|c_{2}\right|^2 \right> + \left|S_{21}\right|^2kT_{0} \label{eqn:Ce_2_sqr} \\
\left< \left|c_{e3}\right|^2 \right>_{2,3} = \left< \left|c_{3}\right|^2 \right> + \left|S_{31}\right|^2kT_{0} \label{eqn:Ce_3_sqr_B} \\
\left< c_{e2}c_{e3}^* \right>_{2,3} = \left< c_{2}c_{3}^* \right> + S_{21}S_{31}^* kT_{0} 
\label{eqn:Ce2_ce3}
\end{eqnarray}
from which we gain knowledge of three further unknowns in~\eqref{eqn:Noise_corr}, $\left< \left|c_{2}\right|^2 \right>$ and $\left< c_{2}c_{3}^* \right>$.

\subsubsection{Terminating Port~3}
\label{sec:term_port3}
The last remaining entry in~\eqref{eqn:Noise_corr}, $\left< c_{2}c_{1}^* \right>$, may be obtained by terminating port~3 and repeating the steps above. It can be shown that 
\begin{eqnarray}
\left< c_{e1}c_{e2}^* \right>_{1,2} = \left< c_{1}c_{2}^* \right> + S_{13}S_{23}^* kT_{0}
\label{eqn:Ce1_ce2}
\end{eqnarray}
For a DISO LNA, it is reasonable to omit this step by assuming that crosstalk between the inputs of the constituent single-ended (SE) amplifiers is negligible ($\left< c_{2}c_{1}^* \right>\approx0$).

\subsection{Measurement Steps for DISO Noise Parameter Extraction}
\label{sec:steps}
In brief, the measurement steps are as follows:
\begin{enumerate}
\item{Measure $S$-parameters.}
\item{Terminate port~2, source pull port~1 $(\Gamma_{s2}=0, \Gamma_{s1})$ using an o.c. long cable or a source tuner. Extract the noise parameters and then convert them to the effective two-port noise waves $\left< \left|c_{e1}\right|^2 \right>_{1,3}, \left< \left|c_{e3}\right|^2 \right>_{1,3}$ and $\left< c_{e1}c_{e3}^* \right>_{1,3}$. Then, using \eqref{eqn:Ce_1_sqr}--\eqref{eqn:Ce1_ce3} obtain $\left< \left|c_{1}\right|^2 \right>, \left< \left|c_{3}\right|^2 \right>, \left< c_{1}c_{3}^{*} \right>$.}
\item{Repeat with measurement 2: terminate port 1, source pull port 2 $(\Gamma_{s1}=0, \Gamma_{s2})$. Obtain $\left< \left|c_{2}\right|^2 \right>$ and $\left< c_{2}c_{3}^* \right>$.}
\item{If necessary, repeat with measurement 3: terminate port~3, source pull port~1 $(\Gamma_{s3}=0, \Gamma_{s1})$. Obtain $\left< c_{1}c_{2}^* \right>$.}
\end{enumerate}

\subsection{Interpreting the Measurement Results for a DISO Device}
\label{sec:apply}
\begin{figure}[htb]
	\begin{center}
		\includegraphics[width=0.7\linewidth]{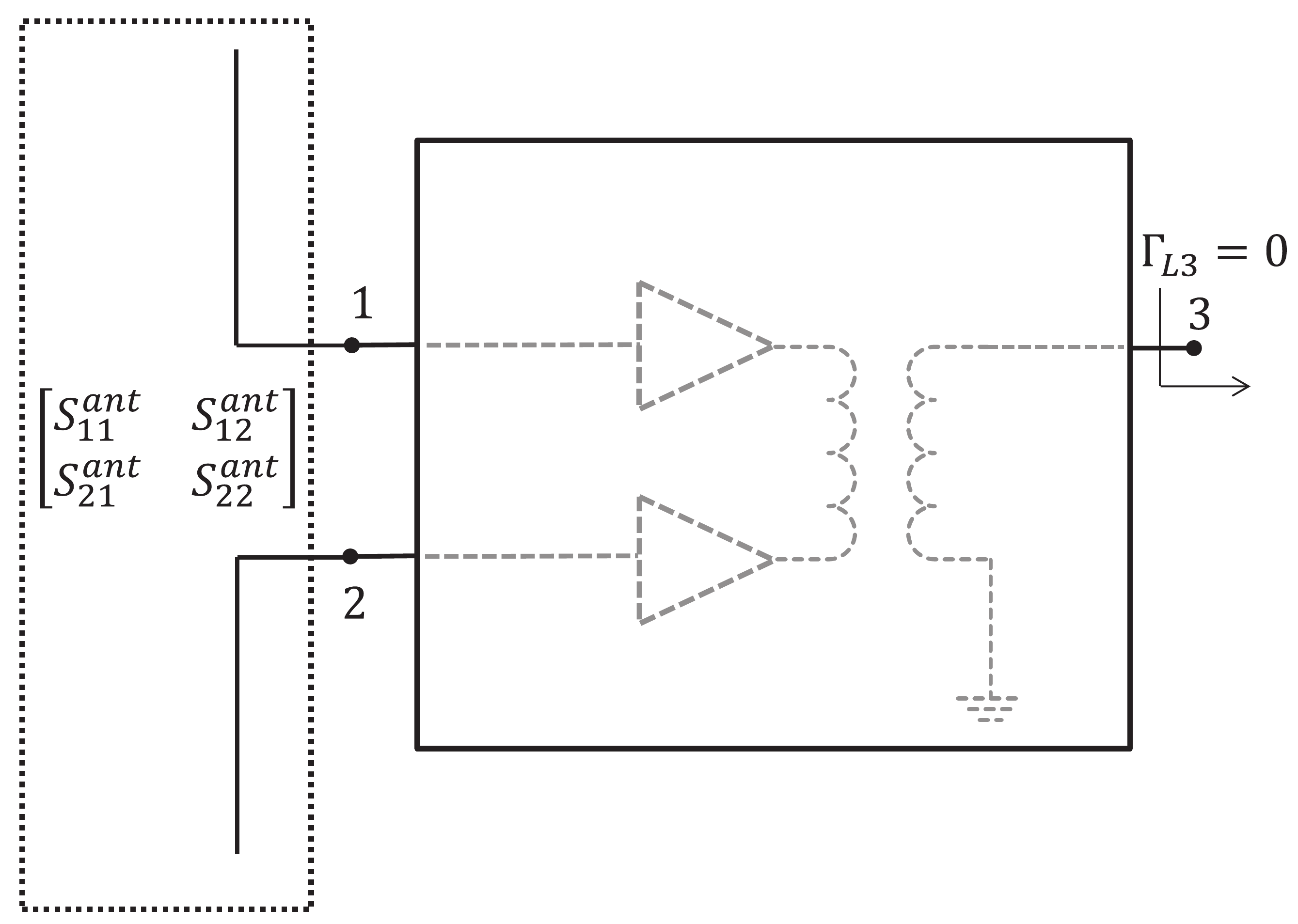}
	\end{center}
	\caption{Differential-input AUT connected to the DISO LNA.}
	\label{fig:Ant_DISO}
\end{figure}

We combine ports~1 and~2 into a differential port~1 (d1). Port~3 becomes the new single-ended port~2 (s2). The noise factor expression becomes
\begin{eqnarray}
&& F_{s2}\left(\Gamma_{s:d1}\right) = 1+ \label{eqn:F_ds} \\
&&\frac{\left[\left(\mathbf{1}-\mathbf{S}_{ds}\mathbf{\Gamma}_{d1}\right)^{-1}\hat{\mathbf{N}}_{ds}\left(\mathbf{1}-\mathbf{S}_{ds}\mathbf{\Gamma}_{d1}\right)^{-1H}\right]_{2,2}}
{kT_{0}\left[\left(\mathbf{1}-\mathbf{S}_{ds}\mathbf{\Gamma}_{d1}\right)^{-1}\mathbf{S}_{ds}\mathbf{A}_{ds}\mathbf{S}_{ds}^{H}\left(\mathbf{1}-\mathbf{S}_{ds}\mathbf{\Gamma}_{d1}\right)^{-1H}\right]_{2,2}} \nonumber
\end{eqnarray}
where 
\begin{eqnarray}
&&\mathbf{S}_{ds} =
\left( \begin{array}{cc}
S_{dd11} & S_{ds12}\\
S_{sd21} & S_{ss22}\\
\end{array} \right)  \label{eqn:S_ds}  \\
&&= 
\left( \begin{array}{cc}
\frac{1}{2}(S_{11}-S_{21}-S_{12}+S_{22}) & \frac{1}{\sqrt{2}}(S_{13}-S_{23})\\
\frac{1}{\sqrt{2}}(S_{31}-S_{32}) & S_{33} 
\end{array} \right) \nonumber
\end{eqnarray}
is the differential to single-ended 2-port $S$~parameters obtained by converting the single-ended $S$-parameters to mixed-mode $S$-parameters~\cite{Fan_03_1271579} and selecting the relevant DISO entries.
\begin{eqnarray}
\mathbf{\Gamma}_{d1}=
\left( \begin{array}{cc}
\Gamma_{s:d1} & 0\\
0 & 0\\
\end{array} \right)
\label{eqn:Gamma_d1}
\end{eqnarray} 
is the differential source reflection coefficient. For a DUT connected to a differential-input antenna, such as the MWA bow-tie antenna, 
\begin{eqnarray}
\Gamma_{s:d1}=\frac{1}{2}\left(S_{11}^{ant}-S_{21}^{ant}-S_{12}^{ant}+S_{22}^{ant}\right)
\label{eqn:Gamma_d1_bal_ant}
\end{eqnarray}
where $S_{ij}^{ant}$ is a single-ended $S$ parameter of the antenna under test (AUT) as shown in the Fig.~\ref{fig:Ant_DISO}. The bottom right entry of $\mathbf{\Gamma}_{d1}$ is zero, which represents matched termination at the output of the DISO device.

\begin{eqnarray}
&&\mathbf{N}_{ds}= \label{eqn:N_ds} \\
&&\left( \begin{array}{cc}
\mathrm{N}_{ds}^{(1,1)}& \frac{1}{\sqrt{2}}\left(\left<c_{1}c_{3}^*\right>-\left<c_{2}c_{3}^*\right>\right)\\
\frac{1}{\sqrt{2}}\left(\left<c_{3}c_{1}^*\right>-\left<c_{3}c_{2}^*\right>\right) & \left<\left|c_{3}\right|^2\right> \\
\end{array} \right) \nonumber
\end{eqnarray}
is the DISO noise correlation matrix (obtained through single-ended to mixed-mode conversion similar to $\mathbf{S}_{ds}$) in which $\mathrm{N}_{ds}^{(1,1)}=\frac{1}{2}\left(\left<\left|c_{1}\right|^2\right>-2\mathrm{Re}\left<c_{1}c_{2}^*\right>+\left<\left|c_{2}\right|^2\right>\right)$. If desired, the noise parameters of the DISO device may be calculated by converting \eqref{eqn:N_ds} to noise parameters. Finally,
\begin{eqnarray}
\mathbf{A}_{ds}=
\left( \begin{array}{cc}
1 & 0 \\
0 & 0 \\
\end{array} \right) 
\label{eqn:A_ds}
\end{eqnarray}
\added{The top left entry represents an ambient temperature differential noise at the input of the DISO device. In our context, this is the noise at the differential AUT port where the AUT is surrounded by thermal noise at ambient temperature.} 

This completes the measurement and interpretation steps. The noise temperature of the DISO device (in this case, the LNA) for arbitrary source reflection coefficient is given by (see Appendix)
\begin{eqnarray}
T_\mathrm{LNA}(\Gamma_{s})=
\frac{\mathrm{N}_{ds}^{(1,1)}\left|\chi\right|^{2}-2\mathrm{Re}\left(\mathrm{N}_{ds}^{(1,2)}\chi\right)+\left<\left|c_{3}\right|^{2}\right>}
{k\left| S_{sd21} \right|^2 \left(1-\left|\Gamma_{s}\right|^2\right)/ \left|1-S_{dd11}\Gamma_{s}\right|^2} 
\label{eqn:T_DISO_gamS}
\end{eqnarray}
where $\chi=S_{sd21}\Gamma_{s}/(S_{dd11}\Gamma_{s}-1)$.

\section{Measurements}
\label{sec:meas_res}

\subsection{Cable Selection}
\label{subsec:Cable}
The expression in \eqref{eqn:T_RX_n} can be transformed into a linear function with four coefficients \cite{Hu_04_1295149}. The relationship of the coefficients to DUT temperature is given by 
\begin{eqnarray}
\label{eqn:abcd_extract}
\mathbf{t}
&=&
\left( \begin{array}{cccc}
1 & \frac{1}{1-\gamma_{1}^{2}} & \frac{\gamma_{1}\cos(\theta_{1})}{1-\gamma_{1}^{2}} & \frac{\gamma_{1}\sin(\theta_{1})}{1-\gamma_{1}^{2}}\\
1 & \frac{1}{1-\gamma_{2}^{2}} & \frac{\gamma_{2}\cos(\theta_{2})}{1-\gamma_{2}^{2}} & \frac{\gamma_{2}\sin(\theta_{2})}{1-\gamma_{2}^{2}}\\
\vdots & \vdots & \vdots & \vdots\\
1 & \frac{1}{1-\gamma_{m}^{2}} & \frac{\gamma_{m}\cos(\theta_{n})}{1-\gamma_{m}^{2}} & \frac{\gamma_{m}\sin(\theta_{m})}{1-\gamma_{m}^{2}}
\end{array} \right)
\mathbf{a}\nonumber \\
&=&\mathbf{X}\mathbf{a}
\end{eqnarray}
where the vector $\mathbf{t} = [T_{1}, T_{2}, \dots, T_{m}]^{T}$ contains the DUT temperature for a given $\Gamma_{n} = \gamma_{n}e^{j\theta_{n}}$ at the input, the $\mathbf{X}$ matrix is the source reflection coefficient matrix which includes contributions from matched ($\gamma_n=0$) and mismatched sources of equal row lengths; $\mathbf{a} = [a, b, c, d]^{T}$ contains the four coefficients. 

The conversion between $\mathbf{a}$ and noise parameters are as follows \cite{Hu_04_1295149}
\begin{eqnarray}
\label{eqn:abcd2noise_para}
\begin{aligned}
\Delta  &=& \sqrt{b^2 - c^2 - d^2} \\
T_{\mathrm{min}}  &=& a + \frac{b+\Delta}{2} \\
N  &=& \frac{\Delta}{4T_{0}} \\
\gamma_{\mathrm{opt}}  &=& \sqrt{\frac{b-\Delta}{b+\Delta}} \\
\theta_{\mathrm{opt}}   &=& \tan^{-1}\left(\frac{-d}{-c}\right)
\end{aligned}
\end{eqnarray}
When using an o.c. cable as the mismatched source, we rely on frequency variation to generate various $\Gamma_{n}$'s. Within this frequency window, we assume that $\mathbf{a}$ is relatively constant. We assign the resulting noise coefficients, $\mathbf{a}$, from the least squares solution to the center frequency ($f_{\mathbf{a}}$) of the window
\begin{eqnarray}
	\label{eqn:freq_abcd}
	f_{\mathbf{a}} = \frac{1}{m}\sum_{i=1}^{m}f_{i}
\end{eqnarray}
where $f_{1}$ is the start frequency and $f_{m}$ is the end frequency.  Moving the window by one frequency step $\Delta$, moves $f_{\mathbf{a}}$ by that same amount. This is the method we use to obtain $\mathbf{a}$ as a function of frequency. The window size is determined as follows \cite{Hu_04_1295149}
\begin{eqnarray}
\label{eqn:win_size}
W_{\mathbf{L}} = 0.5\frac{v_c}{L} ~~(\mathrm{Hz})
\end{eqnarray}
where $v_c$ is the phase velocity of the cable and $L$ is the cable length. The size of $W_{L}$ in \eqref{eqn:win_size} is such that $\Gamma$ completes one full rotation on the Smith Chart.

For the frequency range of  50-350~MHz, window size of a few MHz is reasonable. This translates to cable lengths of $\sim20-30$~m. In addition, the selected cable must produce a well-conditioned matrix $\left(\mathbf{X}^{T}\mathbf{X}\right)$ for the least squares
\begin{eqnarray}
\label{eqn:least_square}
\mathbf{a} = \left(\mathbf{X}^{T}\mathbf{X}\right)^{-1}\mathbf{X}^{T}\mathbf{t}
\end{eqnarray}
where $(.)^{T}$ represents a transpose operation. 

\begin{figure}[htb]
	\begin{center}
		\includegraphics[width=0.9\linewidth]{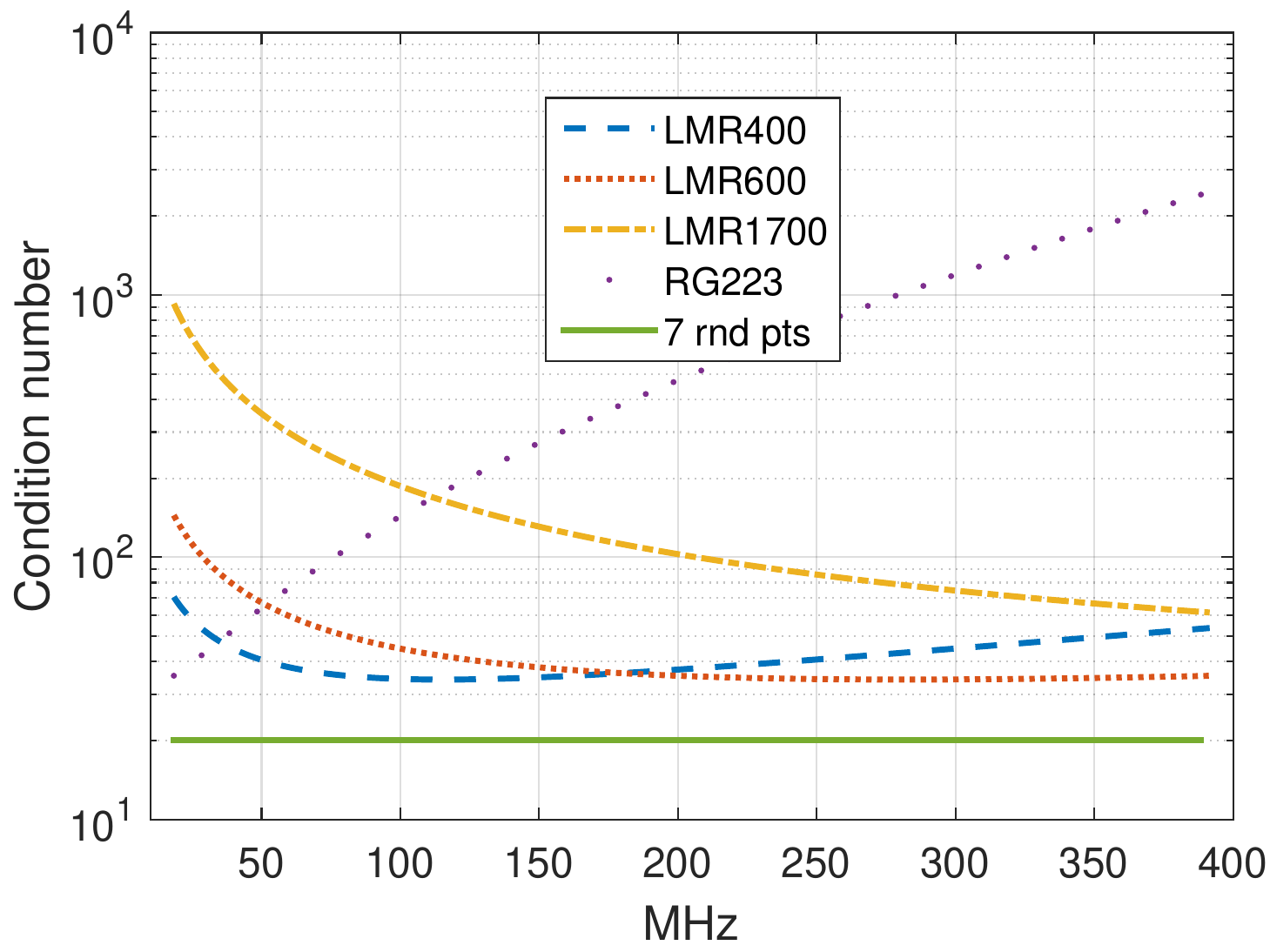}
	\end{center}
	\caption{Predicted condition number based on different types of cable at length of 20~m. The $\Gamma$ of the o.c. cable is given by $e^{-2(\alpha + j\beta)l=e^{-2\alpha l}e^{-j2\pi f/W_{L}}}$. The values for $\alpha$ and $\beta$ can be inferred from the manufacturer's specifications and $l$ is the selected cable length.}
	\label{fig:cond_number}
\end{figure}

Fig. \ref{fig:cond_number} shows the expected condition number for a few example coaxial cables where the cable losses in increasing order are, LMR1700, LMR600, LMR400, RG223. To achieve a condition number below 50, the o.c cable reflection coefficient must be within the bounds $0.69<|\Gamma|<0.89$. To estimate a low condition number for comparison, we obtain the \emph{minimum} condition number from 1000 random trials involving seven random impedances and seven $\gamma=0$ points  (total of 14~points) to form matrix $\mathbf{X}$ at every trial. The result of this experiment is condition number of 20. We select a 20~m LMR400 cable as it provides consistently low condition numbers ($~\sim 30$ to $~\sim 50$) throughout the frequency band of interest (50-350 MHz) and is reasonably close to the outcome of the random trials. 

A low condition number guarantees linear independence in $\mathbf{X}$ and lowers uncertainties in $\mathbf{a}$. Similar work of minimizing uncertainties in measured noise parameters by means of four impedance points selection using an impedance tuner is discussed in \cite{7360952}. Our selected cable with frequency window as given in \eqref{eqn:win_size} satisfies the bounds in $|\Gamma|$. The combination of this cable choice and a matched load samples all of the required regions on the Smith Chart given in~\cite{7360952}.

To further minimize ripples in the extracted parameters, we use a weighting scheme  that places the highest weight at the center frequency which linearly decreases away from the center. This scheme is the triangular weighting scheme proposed in \cite{Hu_04_1295149}
\begin{eqnarray}
	\label{eqn:weighted_abcd}
	\mathbf{t}\circ\mathbf{w} = \left(\mathbf{X}\circ\mathbf{W}\right)\mathbf{a}
\end{eqnarray}
where $\mathbf{w}$ is a column vector containing the weights, $\mathbf{W}$ is a $m \times 4$ matrix formed by duplicating $\mathbf{w}$ four times and ($\circ$) represents an element by element multiplication (Hadamard Product). The window size that we use under this scheme is
\begin{eqnarray}
\label{eqn:win_size_weight}
W_{\mathbf{L}}^{'} = 2.66W_{\mathbf{L}}
\end{eqnarray}
as it results in smoother $\mathbf{a}(f)$ when compared to the proposed factor of $2$ in \cite{Hu_04_1295149}.

\subsection{Noise Receiver Calibration}
\label{subsec:NR_cal}
We turn on the RX's pre-amplifier (Keysight N9030-90017~\cite{Keysight_PXA}) to achieve the lowest noise figure. Based on our selected cable, we expect ripples in $P_{\mathrm{disp}}$ with period of $W_{\mathbf{L}} = 6.3$~MHz. Therefore, we select resolution bandwidth (RBW) of 100~kHz to resolve the ripple. To reduce uncertainty by a factor of 50 in the measurement of $P_{\mathrm{disp}}$, we average 2500 samples for each power measurement.

\begin{figure}[htb]
	\begin{center}
		\includegraphics[width=0.9\linewidth]{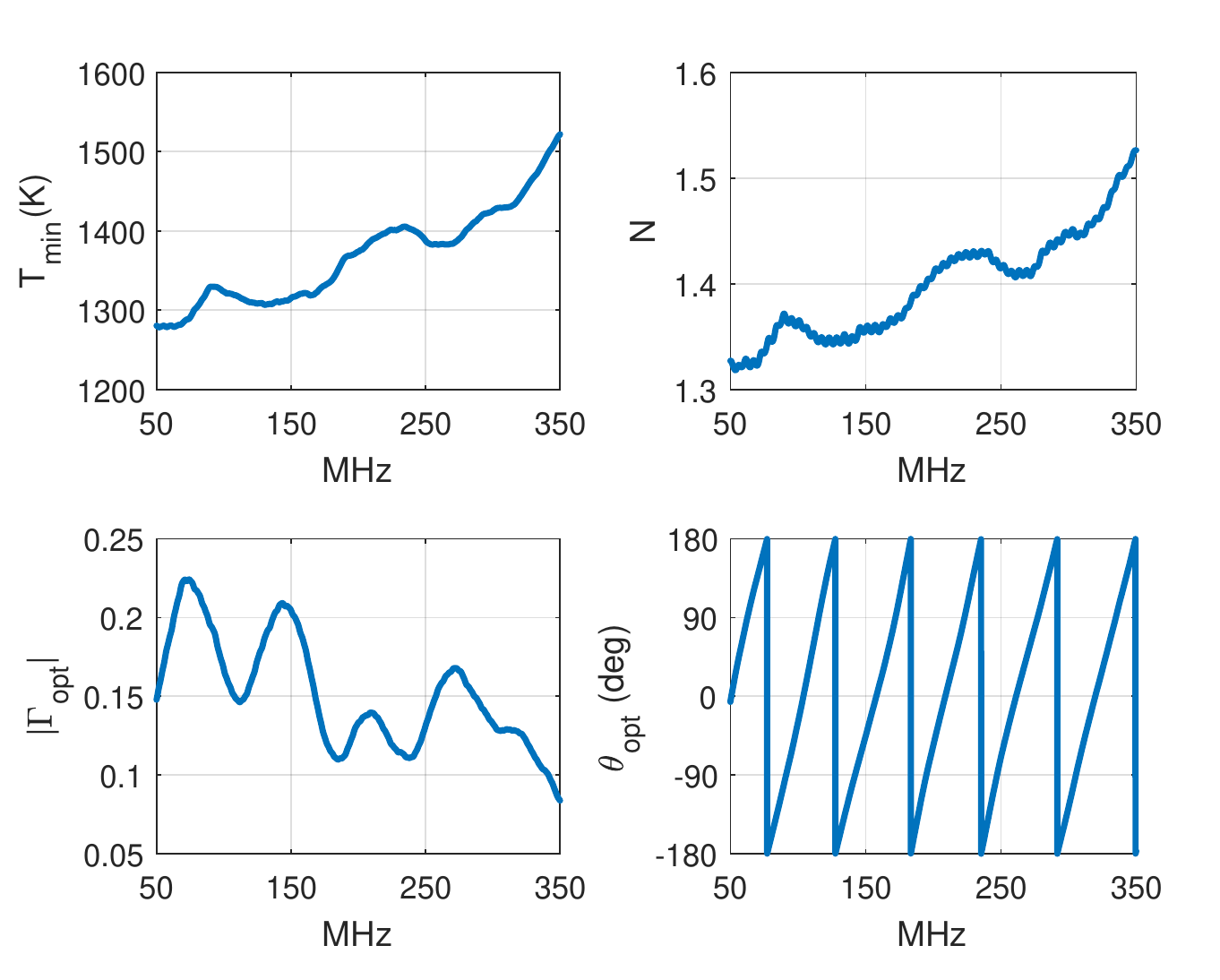}
	\end{center}
	\caption{Extracted noise parameter of the RX. Equation \eqref{eqn:T_LNA_ant} shows that for a DUT with 20 dB gain (as we expect for the MWA LNA), the contribution of $T_{RX}$ to the cascaded noise is reduced by a factor of 100.}
	\label{fig:T_RX_noise_par}
\end{figure}

To calibrate the RX, we use a 15~dB ENR noise source (Agilent 346B) which is a device with the highest ENR in our lab. A high ENR results in less uncertainty due to higher contrast between the displayed power during the hot/cold measurement. Following the steps in \ref{sec:RX cal}, the extracted noise parameters of the RX are shown in Fig. \ref{fig:T_RX_noise_par}.

\subsection{Direct Measurement Results}
\label{subsec:Direct_results}

\begin{figure}[htb]
	
	\begin{center}
		\includegraphics[width=0.9\linewidth]{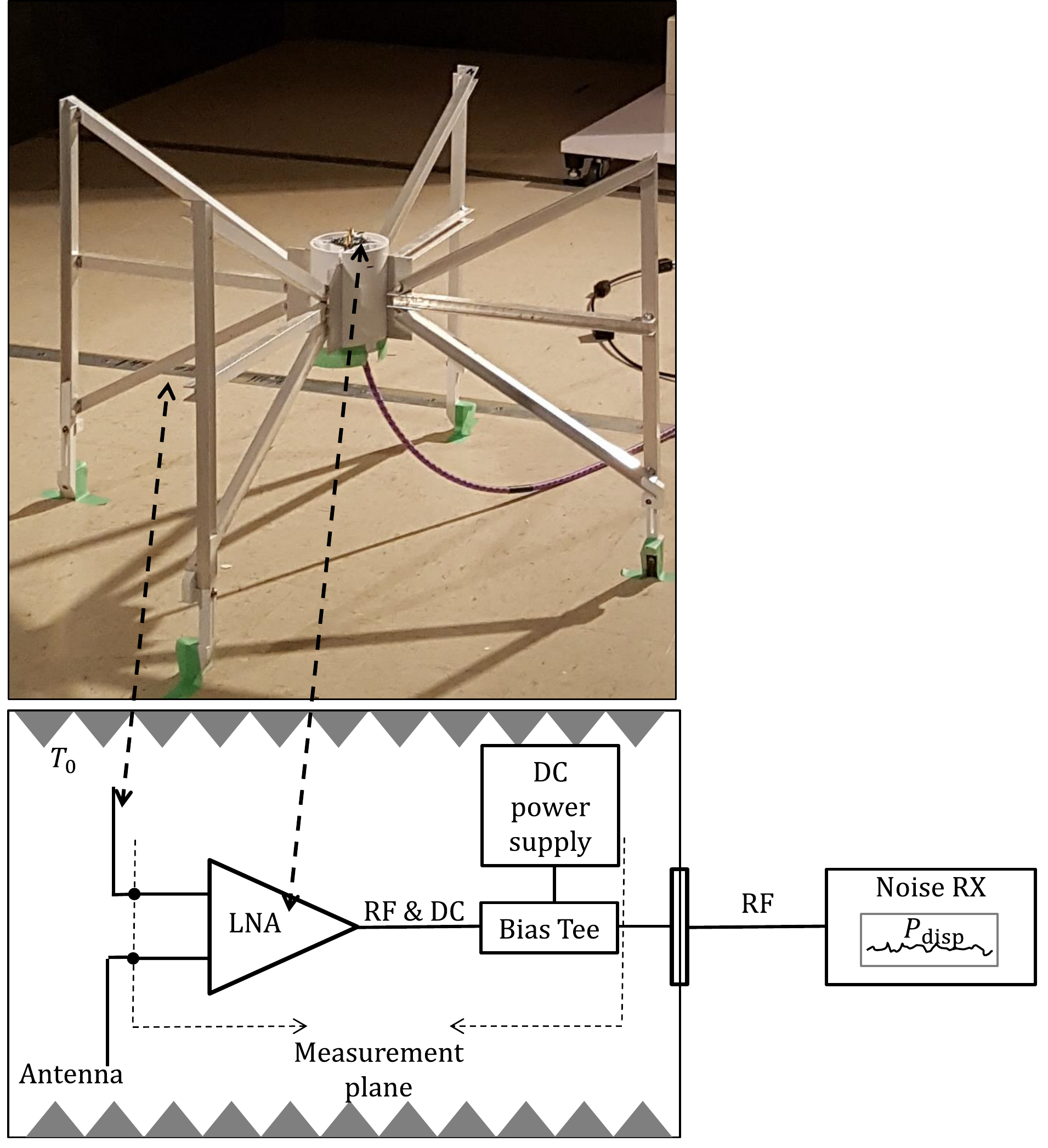}
	\end{center}
	\caption{Measurement setup for direct noise measurement. The antenna, LNA, DC power supply and Bias Tee are located inside the anechoic chamber. The RX and vector network analyzer (VNA) are located outside to prevent interference. The measurement plane shown defines the DUT and any cables/bulkhead between the DUT and the RX are considered to be an extension of the RX input port. Our chamber is semi-anechoic (with metallic ground plane which is compatible with the MWA placement above a metallic ground mesh in the field~\cite{2013PASA...30....7T}) with dimensions approximately $4.9 \times 3.1 \times 2.4$~m rated for 26 MHz to 18 GHz \protect\cite{ETS_FACT3}.}
	\label{fig:measurement_setup}
\end{figure}

Fig. \ref{fig:measurement_setup} shows the measurement setup of the DUT and test equipments for direct noise measurement. Although not shown, the vector network analyzer (VNA) follows a similar setup with three cables entering the anechoic chamber. The DUT is an MWA LNA (DISO device) which has been allowed to reached thermal equilibrium prior to measurement. 
\begin{figure}[htb]
	\begin{center}
		\includegraphics[width=0.9\linewidth]{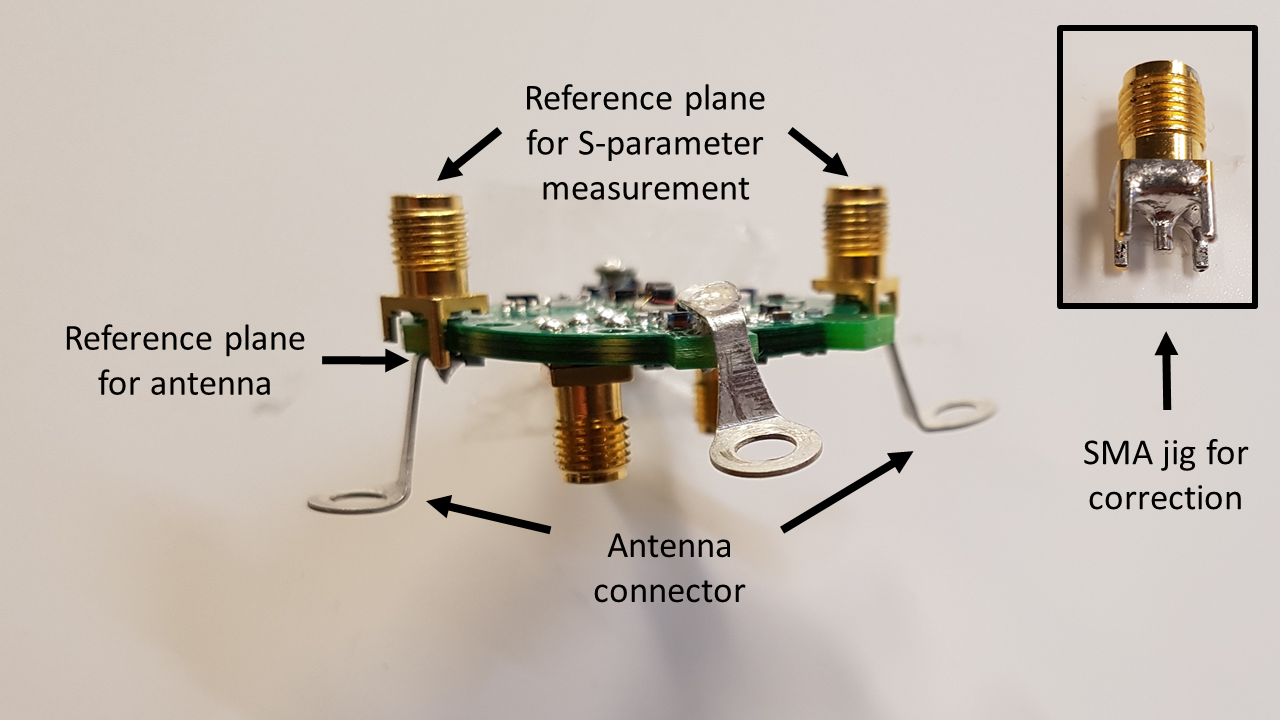}
	\end{center}
	\caption{Side view of LNA. The circuit board was modified to include the SMA connectors at the input ports for $S$ parameter measurements. During power measurement, the LNA was attached to the MWA antenna using the antenna connectors. An SMA jig (short-circuit) was created to estimate the time delay and it was found to be approximately 38~picoseconds ($\approx2.7^{\circ}$ at 200~MHz).}
	\label{fig:EDA_LNA}
\end{figure}

Fig. \ref{fig:EDA_LNA} shows that the reference plane for $S$-parameter measurement is the SMA connectors while the antenna is connected to the connectors soldered to to bottom of the circuit board. We found that a slight delay of 38~ps which translates to a phase shift of $\approx2.7^{\circ}$ at 200~MHz is, in fact, not negligible to the direct measurement. The SMA connectors have de-embedded from the final result.

\subsection{DISO Noise Parameter}
\label{subsec:DISO_noise_par}
\begin{figure}[htb]
	\begin{center}
		\includegraphics[width=0.9\linewidth]{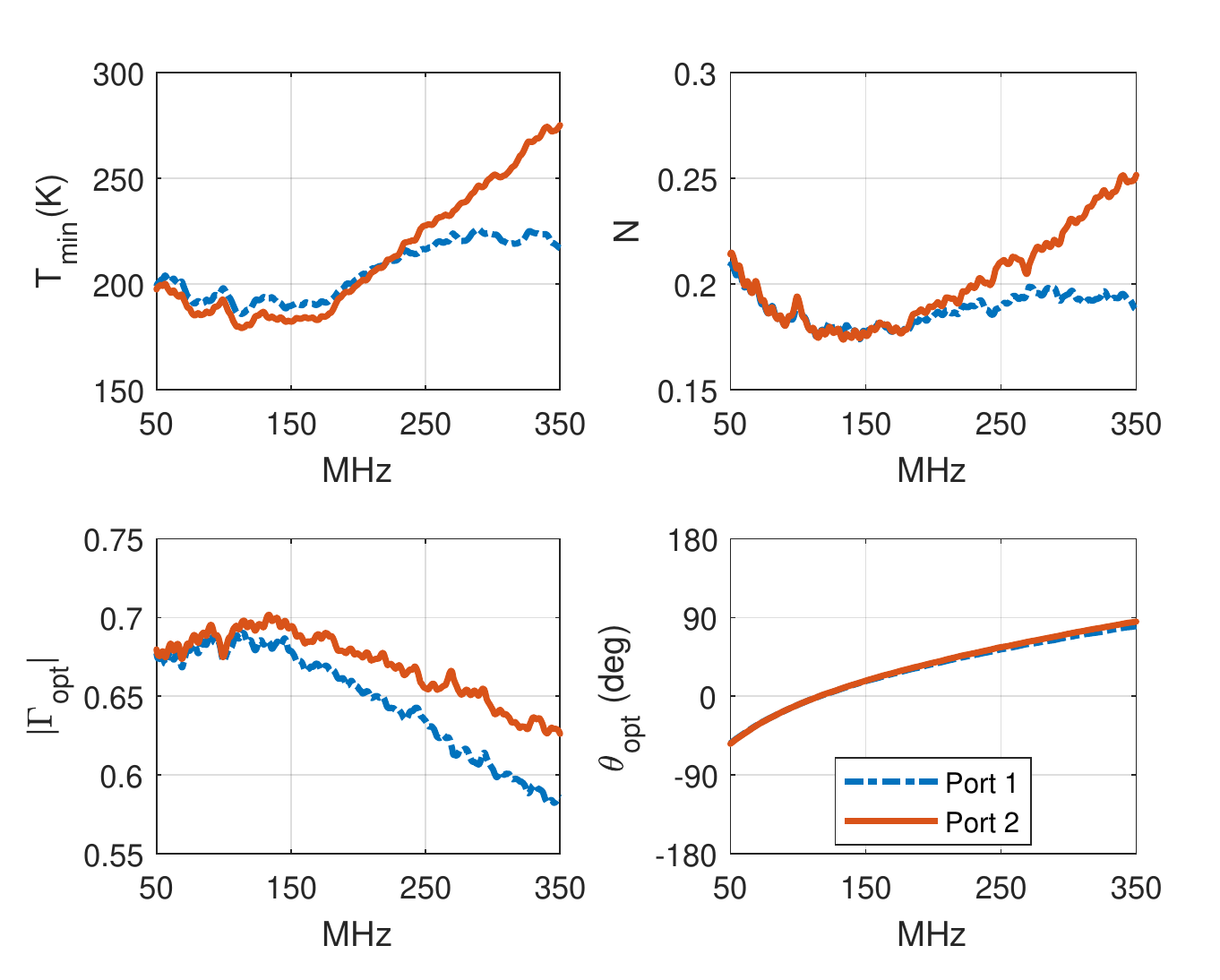}
	\end{center}
	\caption{The extracted noise parameters of a SISO equivalent circuit formed by terminating one of the input port of the LNA with a match load.}
	\label{fig:SISO_noise}
\end{figure}

We follow the steps in Sec.~\ref{sec:steps}. Figure \ref{fig:SISO_noise} shows the extracted noise parameters of the two SISO devices where we terminate the unused input port with a matched load. The measured $T_{\mathrm{min}}$, $N$ and $|\Gamma_{\mathrm{opt}}|$ are reasonable and physically possible throughout. Following this verification, we proceed by combining the results to form a DISO device as discussed in Sec.~\ref{sec:apply}.   

\begin{figure}[htb]
	\begin{center}
		\includegraphics[width=0.9\linewidth]{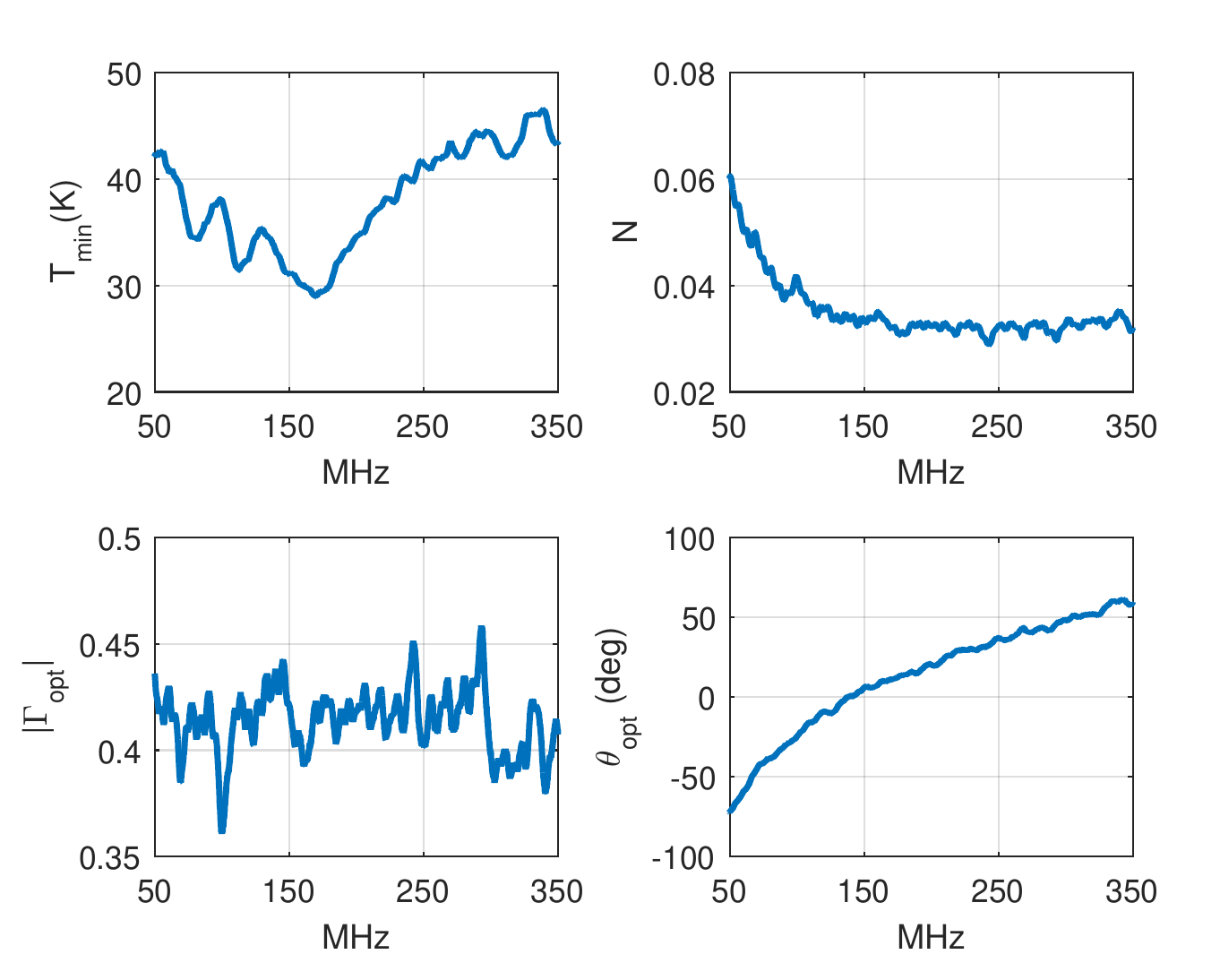}
	\end{center}
	\caption{De-embedded DISO noise parameters of an MWA LNA. The de-embedding assumes that the SMA connector is lossless. We verified that this loss ($\sim1.5$~K) is negligible.}
	\label{fig:DISO_noise_par}
\end{figure}

\begin{figure}[htb]
	\begin{center}
		\includegraphics[width=0.9\linewidth]{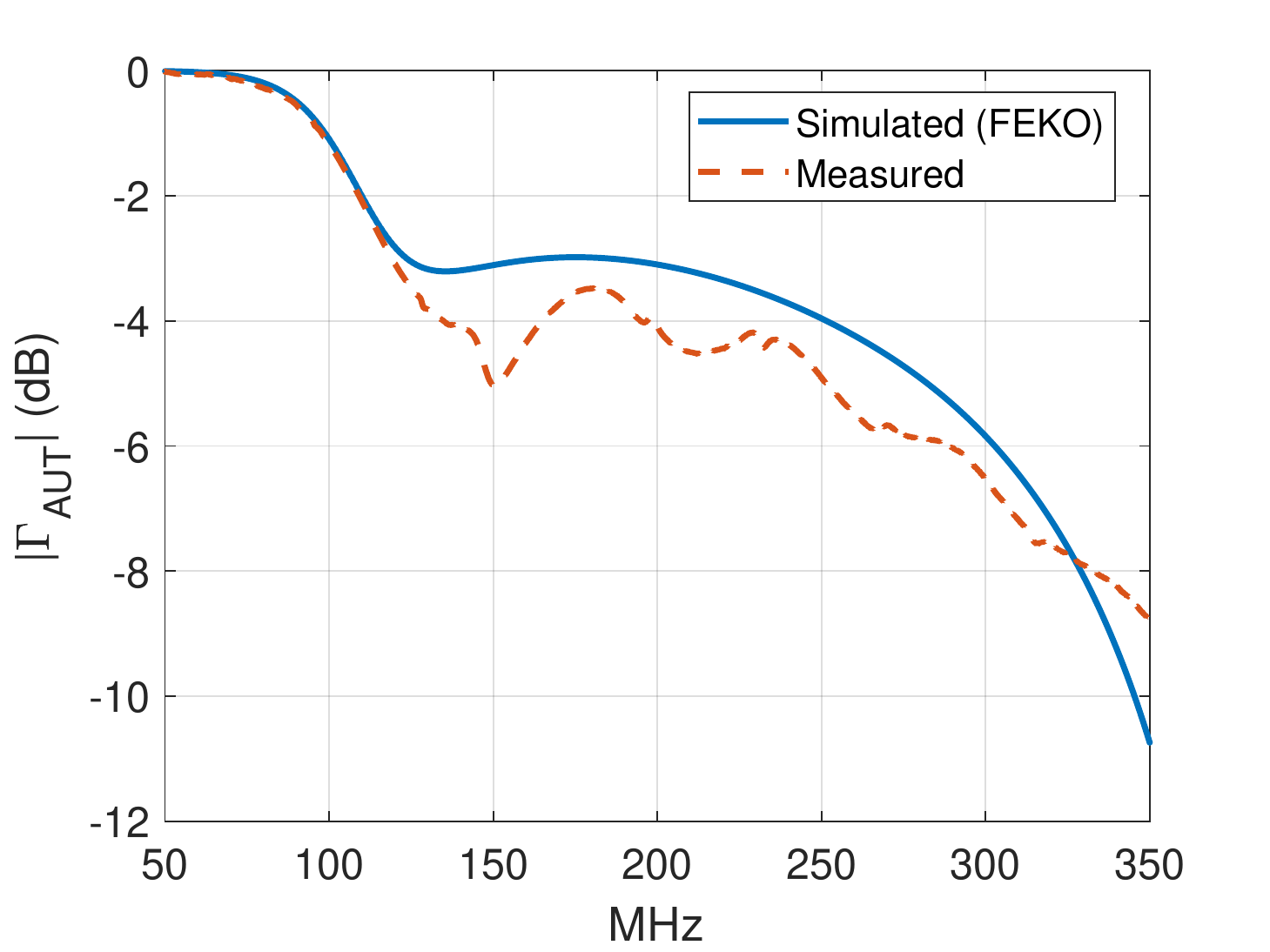}
	\end{center}
	\caption{$|\Gamma_{\mathrm{AUT}}|$ measured in the anechoic chamber and simulated using FEKO.}
	\label{fig:Gamma_AUT}
\end{figure}

Fig. \ref{fig:DISO_noise_par} shows the DISO noise parameters obtained by using \eqref{eqn:S_ds} and \eqref{eqn:N_ds} to convert noise waves back into noise parameters. We can now predict $T_{\mathrm{LNA}}$ for arbitrary $\Gamma_{s}$. To validate $T_{\mathrm{LNA}}$ obtained from the direct measurement method, we measure $\Gamma_{\mathrm{AUT}}$ of the MWA antenna in the anechoic chamber and apply \eqref{eqn:T_DISO_gamS}. The measured $|\Gamma_{\mathrm{AUT}}|$ is reported in Fig.~\ref{fig:Gamma_AUT}. Alternatively, \eqref{eqn:T_RX_n} could be used with the extracted DISO noise parameters (Fig. \ref{fig:DISO_noise_par}).

\begin{figure}[htb]
	\begin{center}
		\includegraphics[width=0.9\linewidth]{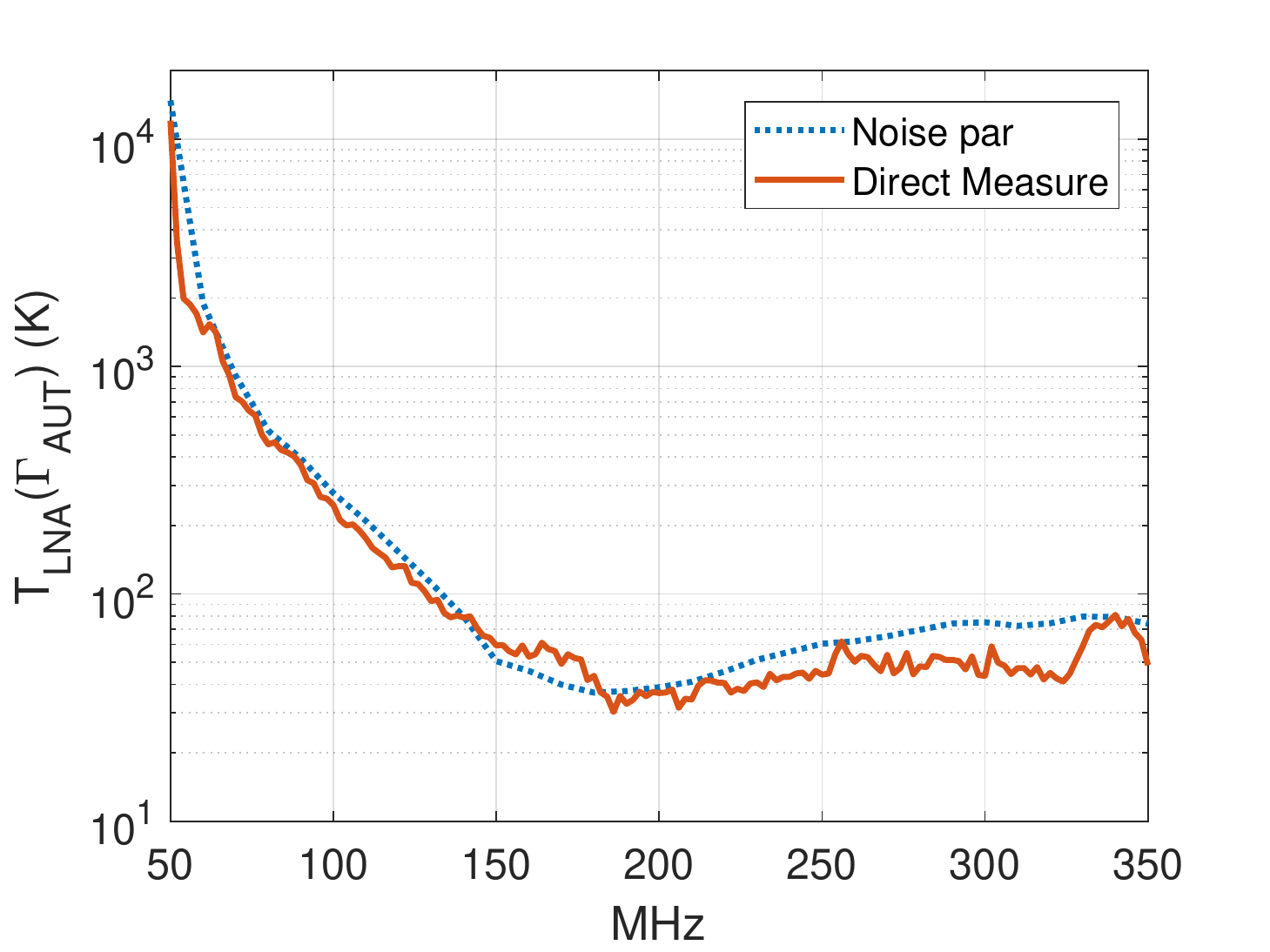}
	\end{center}
	\caption{Comparison between direct measurement and predicted LNA noise temperature based on extracted DISO noise parameters. The ``Noise par'' represents the extracted noise parameters.}
	\label{fig:T_comparison}
\end{figure}

Figure \ref{fig:T_comparison} shows the comparison of the LNA temperature ($\Gamma_{s} = \Gamma_{\mathrm{AUT}}$) obtained using two different methods measured at different times in different locations with independent calibration. It can be seen that overall $T_{\mathrm{LNA}}^{\mathrm{NP}}$ and $T_{\mathrm{LNA}}^{\mathrm{DM}}$ converge to each other which verifies the theory presented in Sec. \ref{sec:extract}. 

\section{Uncertainty Estimates}
\label{sec:uncertainty}

We perform Monte Carlo simulations to estimate the uncertainties in our measurements. The summary of the Monte Carlo simulation steps are as follows: 

\subsection{Direct Measure Monte Carlo Analysis}
\begin{enumerate}
	\item We take the resulting $T_{\mathrm{LNA}}$ based on DISO noise parameters as correct and apply~\eqref{eqn:direct_meas_eqns} to simulate measured $P_{\mathrm{disp}}$ data. We add Gaussian noise with a standard deviation that matches the measured data. 
	\item Using Vector Network Analyzer Uncertainty Calculator provided by Keysight~\cite{VNA_error_calc}, we obtain the uncertainty in our $S$-parameter measurements. The uncertainty calculator takes into account the model of the VNA, type of calibration kit, resolution bandwidth, number of averaging, and source power. The uncertainties depends on the reflection and transmission coefficient of the DUT. We apply the appropriate values to all measured $S$-parameters.
	\item \added{We perturb the $T_{H}$ at every trial based on the uncertainty reported on the certificate of calibration for our Agilent 346B ENR device ($\approx 0.15$~dB in our frequency range) which we treat as an offset across frequency}.   
	\item Using the simulated data and perturbed $S$-parameters, we reprocess the data to obtain $T_{\mathrm{LNA}}$. This is repeated for \added{10000} trials. 
	\item Any unphysical outlier, i.e., $|\Gamma_{\mathrm{AUT}}|>1$ is discarded. 
	\item We calculate the resulting mean and standard deviation. The computed standard deviation is our uncertainty estimate.
\end{enumerate}

\subsection{DISO Monte Carlo Analysis}
\begin{enumerate}
	\item Apply similar steps as 1 to 3 described above to simulate the measurement of each SISO equivalent device. 
	\item We extract $\mathbf{a}$ using the simulated $P_{\mathrm{disp}}$ for each SISO equivalent device. 
	\item We convert $\mathbf{a}$ into noise waves and combine them to form a DISO device. We discard unphysical noise waves and noise parameter outliers, e.g., negative average noise wave power, negative $T_{min}$ and/or negative $N$.
	\item We use the noise parameters of the DISO device to compute $T_{\mathrm{LNA}}$ attached to the AUT. This is repeated for 10000 trials.
	\item We calculate the resulting mean and standard deviation. 
\end{enumerate}

\begin{figure}[htb]
	\begin{center}
		\includegraphics[width=0.9\linewidth]{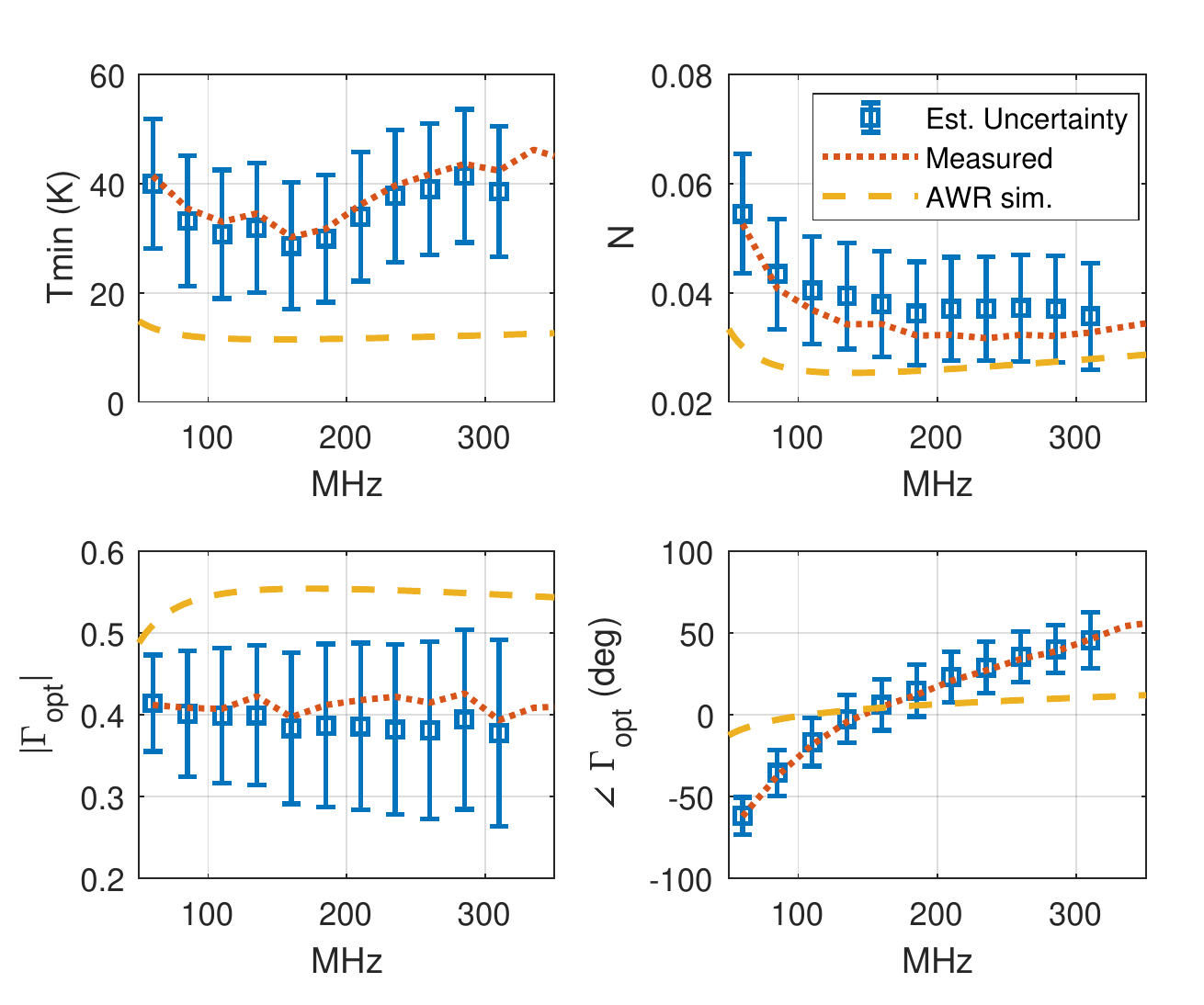}
	\end{center}
	\caption{Comparison of extracted DISO noise parameters of an MWA LNA with estimated uncertainties from Monte Carlo simulation. Also shown are noise parameters extrapolated by NI AWR Microwave Office based on noise parameters supplied by the manufacturer at 500~MHz.}
	\label{fig:DISO_noise_par_error}
\end{figure}

Fig.~\ref{fig:DISO_noise_par_error} reports the measured DISO noise parameters, the result of the Monte Carlo simulation as well as noise parameters based on NI AWR simulation. The simulation model of the MWA LNA in NI AWR is based on Broadcom ATF-54143 LNAs operating at the nominal bias point of $V_{ds} =3$~V and $I_{ds}=60$~mA. The simulation relies on $S$-parameters and noise parameters provided by the data sheet~\cite{ATF-54143}. However, the noise parameters in the frequency range of interest are not available; the lowest frequency available is 500~MHz. AWR extrapolates the noise parameters to 50-350 MHz which may be acceptable as a starting point in engineering design. Fig.~\ref{fig:DISO_noise_par_error} shows that the measurement is generally well within the uncertainty estimates of the Monte Carlo simulation. However, the AWR extrapolated data are outside of the uncertainty estimates. Most notably, the AWR $T_{\mathrm{min}}$ is consistently below the lower uncertainty estimates. 

\begin{figure}[htb]
	\begin{center}
		\includegraphics[width=0.9\linewidth]{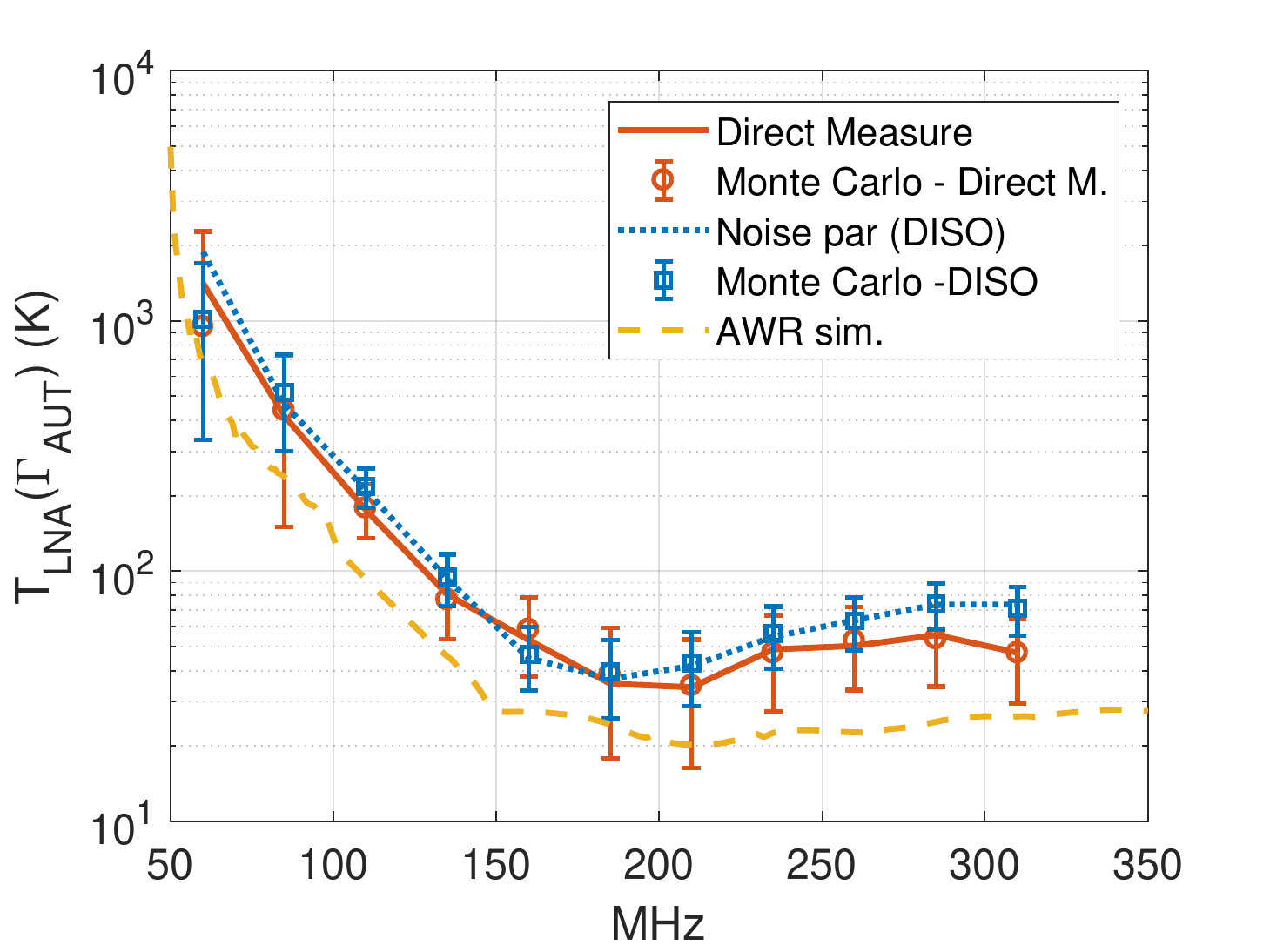}
	\end{center}
	\caption{Comparison of $T_{\mathrm{LNA}}(\Gamma_{\mathrm{AUT}})$ based on measured DISO noise parameters, direct measurement, the result of the Monte Carlo simulation for both methods and NI AWR.}
	\label{fig:T_comparison_error}
\end{figure}

Fig.~\ref{fig:T_comparison_error} shows the resulting $T_{\mathrm{LNA}}(\Gamma_{\mathrm{AUT}})$ based on the measurements, Monte Carlo simulation, and NI AWR simulation. We note the relative convergence of the measurements and the Monte Carlo simulations. The uncertainties for the direct measurement are \added{slightly} larger than the DISO method. This is summarized at the spot frequencies reported in Tab.~\ref{tab:rel_error}. Also notable is underestimation by a factor of $\sim 2-3$ of the calculated noise temperature based on extrapolated noise parameters, which suggests that it is only a first-order estimate. 

To identify the dominant uncertainty contributor, we run the Monte Carlo simulations with only one type uncertainty at a time. The results suggest that at low frequencies, the dominant contributor is $\Gamma_{\mathrm{AUT}}$ for both methods. This is fully expected as $|\Gamma_{\mathrm{AUT}}|$ approaches unity at low frequencies. At high frequencies, the dominant uncertainty contributor for the direct measurement is the uncertainty in ENR \added{and the $S$-parameter uncertainty of the DUT}, whereas for the DISO case, it is uncertainty in the \added{ENR.} 

\added{To gain insight into the interplay of the dominant uncertainty contributors in our measurement (which are $\Gamma_s$, $S$-parameters of the DUT, and the ENR), we take \eqref{eqn:direct_meas_eqns} to be error-free and substitute them into \eqref{eqn:T_LNA_ant} where $G_{A}$ and $G_{P}^{RX}B$ contain uncertainty. After simplification, it can be shown that}
\begin{eqnarray}\added{
\left|\Delta T_{\mathrm{LNA}}(\Gamma_{s})\right| \approx \left|\frac{G_{A}G_{P}^{RX}B}{\tilde{G}_{A}\tilde{G}_{P}^{RX}\tilde{B}}-1\right|\left[T_{\mathrm{LNA}}(\Gamma_{s})+T_0\right]
\label{eqn:T_direct_trend}}
\end{eqnarray}
\added{where the quantities with $\tilde{.}$ are dominated by uncertainties due to $\Gamma_s$ and  $S$-parameters of the DUT in the case of $\tilde{G}_{A}$,  and uncertainties due to ENR in the case of $\tilde{G}_{P}^{RX}\tilde{B}$. Uncertainty in the ENR in \% translates directly to uncertainty in $\tilde{G}_{P}^{RX}\tilde{B}$ in \%. Therefore, in the DISO method at high frequencies where $T_{\mathrm{LNA}}(\Gamma_s) \ll T_0$, $\left|\Delta T_{\mathrm{LNA}}(\Gamma_{s})\right|\approx |\Delta ENR\,(\%)|T_0$. This ultimately sets the limit for the lowest measurable $T_{\mathrm{LNA}}$}. \added{ For the direct measurement, the uncertainty in $\tilde{G}_A$ is approximately $3\%$ at high frequencies which is comparable to the ENR uncertainty. Their contributions are equally dominant which explains the slight increase of the overall uncertainty.}

\begin{table}
	\centering
	\begin{tabular}{|c|c|c|c|}
		\hline
		$f_{\mathrm{MHz}}$ & $T_{\mathrm{LNA}}^{\mathrm{NP}}\pm\Delta_{\mathrm{MC}}$~(K) & $T_{\mathrm{LNA}}^{\mathrm{DM}}\pm\Delta_{\mathrm{MC}}$~(K)  & $T_{\mathrm{LNA}}^{\mathrm{AWR}}$~(K) \\ [0.5ex]
		\hline \hline
		60	& 1886$\pm$680 & 1413$\pm$1331 & 690 \\
		85	& 463$\pm$217 & 416$\pm$288 & 237 \\
		110	&  211$\pm$39 & 175$\pm$43 & 93\\
		135	& 93$\pm$22 & 81$\pm$23 & 46\\
		160	& 45$\pm$13 & 53$\pm$20 & 27\\
		185	& 37$\pm$14 & 35$\pm$21 & 24\\
		210	& 42$\pm$14 & 34$\pm$18 & 20\\
		235	& 55$\pm$16 & 49$\pm$20 & 23\\
		260	& 64$\pm$15 & 50$\pm$19 & 23\\
		285	& 73$\pm$16 & 56$\pm$19 & 25\\
		310	& 73$\pm$16 & 47$\pm$17& 26\\
		\hline
	\end{tabular}
	\caption{Reported $T_{\mathrm{LNA}}$ at spot frequencies obtained using three different methods. ($.^{NP}$), ($.^{DM}$) and ($.^{AWR}$) represents, noise parameters (de-embedded), direct measurement and AWR (simulation with extrapolated noise parameters from 500~MHz) method respectively. The estimated uncertainties obtained from Monte Carlo simulations are shown as $\pm\Delta_{\mathrm{MC}}$.}
	\label{tab:rel_error}
\end{table}

\section{Conclusion}
\label{sec:concl}
We presented simple methods of measuring the noise temperature of a DISO LNA connected to an antenna at low frequencies (50-350 MHz). The direct method uses an anechoic chamber as an ambient noise source. The noise parameter extraction method combines the noise parameters from two SISO measurements obtained by terminating one unused input port. The DISO noise parameters are reconstructed from these results.

The convergence of the measurement results using the MWA DISO LNA and antenna suggests that both methods are reliable. \added{Monte Carlo simulation suggests that the uncertainty of noise parameter extraction method is slightly better than the direct measurement and is limited by the uncertainty of the ENR device}. The noise parameter extraction method we present is a low-cost option using a long open circuit cable; however, the theory and measurement steps outlined here are fully applicable to any impedance tuners. Furthermore, we show that the DISO noise parameter extraction can be performed with one source tuner at a time which saves significant cost.

Finally, we confirmed that the practice of using extrapolated noise parameters based on data sheet values is only a first-order estimate. For the case of the MWA LNA, the simulation using extrapolated noise parameters from 500~MHz leads to noise underestimation of approximately a factor of two.

\section*{Acknowledgment}
This scientific work makes use of the Murchison Radio-astronomy Observatory, operated by CSIRO. We acknowledge the Wajarri Yamatji people as the traditional owners of the Observatory site. Support for the operation of the MWA is provided by the Australian Government (NCRIS), under a contract to Curtin University administered by Astronomy Australia Limited. We acknowledge the Pawsey Supercomputing Centre which is supported by the Western Australian and Australian Governments.

The authors thank Dr. Marcin Sokolowski, A/Prof.~Randall Wayth and A/Prof.~Cathryn Trott for useful discussions on this topic. We acknowledge Mr. Dave Kenney and Mr. Jon Tickner for providing laboratory and prototyping support.

\section*{Appendix}
\label{sec:app}
\added{The purpose for this Appendix is to explain the emergence of \eqref{eqn:T_DISO_gamS} from \eqref{eqn:F_ds}. In particular, we explain the need for the extra factor $(1-|\Gamma_s|^2)$ in the numerator of \eqref{eqn:T_DISO_gamS}}. For a two-port device (port~1 input, port~2 output), \eqref{eqn:Multiport_F} becomes
\begin{eqnarray}
&&F(\Gamma_{s})= 1+ \nonumber \\
&&+\frac{\left<\left|c_{1}\right|^{2}\right>\left|\chi\right|^{2}-2\mathrm{Re}\left(\left<c_{1}c_{2}^{*}\right>\chi\right)+\left<\left|c_{2}\right|^{2}\right>}
{kT_{0}\left(\left|S_{11}\right|^{2}\left|\chi \right|^{2}-2\mathrm{Re}\left(S_{11}S_{21}^{*}\chi\right)+\left|S_{21}\right|^{2}\right)} \nonumber \\
&& = 1 +
\frac{\left<\left|c_{1}\right|^{2}\right>\left|\chi\right|^{2}-2\mathrm{Re}\left(\left<c_{1}c_{2}^{*}\right>\chi\right)+\left<\left|c_{2}\right|^{2}\right>}
{kT_{0}\left(\left| S_{21} \right|^2 / \left|1-S_{11}\Gamma_{S}\right|^2 \right)} 
\label{eqn:F_2port_gamS}
\end{eqnarray}
where 
\begin{eqnarray}
\chi=\frac{S_{21}\Gamma_{s}}{S_{11}\Gamma_{s}-1}
\label{eqn:chi}
\end{eqnarray}
\added{Equation \eqref{eqn:F_2port_gamS} can also be derived by following a signal flow diagram of a two-port $S$-parameter network driven with a source impedance $Z_s$ with load reflection coefficient $\Gamma_L=0$~\cite{Gonzalez_1997_ch2}. The correction factor is needed in the denominator of the second term in \eqref{eqn:F_2port_gamS} because}  
\begin{eqnarray}
kT_{0}\left(\left| S_{21} \right|^2 / \left|1-S_{11}\Gamma_{S}\right|^2 \right) 
\label{eqn:denom_F}
\end{eqnarray}
\added{is the power delivered to $Z_L=Z_0$ \emph{given that} noise power density of $kT_{0}$ is \emph{delivered} to $Z_{in}=Z_0$ at the input. Therefore, to bring \eqref{eqn:F_2port_gamS} in line with the $F(\Gamma_{s})$ formulas based on noise parameters~\cite{Hu_04_1295149, Leo_16_7506352}, we need to modify \eqref{eqn:denom_F} such that $kT_{0}$ is the \emph{power available} at the source. A source with available noise power density of $kT_{0}$ with source reflection coefficient $\Gamma_s$ delivers $kT_{0}(1-|\Gamma_s|^2)$ noise power density to $Z_{in}=Z_0$. With this correction, \eqref{eqn:F_2port_gamS} becomes}
\begin{eqnarray}
&&F(\Gamma_{s})= \nonumber \\
&&  1 +
\frac{\left<\left|c_{1}\right|^{2}\right>\left|\chi\right|^{2}-2\mathrm{Re}\left(\left<c_{1}c_{2}^{*}\right>\chi\right)+\left<\left|c_{2}\right|^{2}\right>}
{kT_{0}(1-|\Gamma_s|^2)\left(\left| S_{21} \right|^2 / \left|1-S_{11}\Gamma_{S}\right|^2 \right)} 
\label{eqn:F_2port_gamS_corr}
\end{eqnarray}


\end{document}